\documentclass[aps,prl,reprint,superscriptaddress,longbibliography]{revtex4-2}

\usepackage[utf8]{inputenc}
\usepackage{amsmath}
\usepackage{amssymb}
\usepackage{mathtools}
\usepackage[normalem]{ulem}
\usepackage{xcolor}
\usepackage{siunitx}
\usepackage{algorithm}
\usepackage[noend]{algpseudocode}
\usepackage[multiple]{footmisc}
\usepackage{booktabs}
\usepackage{multirow}
\usepackage{setspace}

\usepackage{geometry}

\definecolor{linkColor}{RGB}{0,80,150}
\usepackage{hyperref}
\hypersetup{
    colorlinks=true,
    allcolors=linkColor,
    pdfborder={0 0 0},
    pdfencoding = auto
}

\definecolor{myGreen}{RGB}{19,132,23}

\def\cT{c_\mathrm{T}}
\def\cM{c_\mathrm{M}}

\def\cMqss{\tilde{c}_\mathrm{M}}

\def\rhoM{\rho_\mathrm{M}}
\def\rhoT{\rho_\mathrm{T}}
\def\barRhoT{\bar{\rho}_{\mathrm{T}}}
\def\barRhoM{\bar{\rho}_{\mathrm{M}}}

\def\kon{k_\mathrm{on}}
\def\koff{k_{\mathrm{off}}}
\def\vm{v_\mathrm{m}}
\def\DT{D_\mathrm{T}}
\def\DM{D_\mathrm{M}}

\def\vg{v_\mathrm{g}}
\def\vs{v_\mathrm{s}}

\def\vsqss{\tilde{v}_\mathrm{s}}

\begin{document}
\title{
Length Regulation Drives Self-Organization in Filament-Motor Mixtures
}
\author{Moritz Striebel}
\thanks{M.S. and F.B. contributed equally to this work.}
\affiliation{
Arnold Sommerfeld Center for Theoretical Physics and Center for NanoScience, Department of Physics, Ludwig-Maximilians-Universit\"at M\"unchen, Theresienstrasse 37, D-80333 Munich, Germany
}
\thanks{}
\author{Fridtjof Brauns}
\thanks{M.S. and F.B. contributed equally to this work.}
\affiliation{
Arnold Sommerfeld Center for Theoretical Physics and Center for NanoScience, Department of Physics, Ludwig-Maximilians-Universit\"at M\"unchen, Theresienstrasse 37, D-80333 Munich, Germany
}
\author{Erwin Frey}
\affiliation{
Arnold Sommerfeld Center for Theoretical Physics and Center for NanoScience, Department of Physics, Ludwig-Maximilians-Universit\"at M\"unchen, Theresienstrasse 37, D-80333 Munich, Germany
}
\affiliation{
 Max Planck School Matter to Life, Hofgartenstraße 8, D-80539 Munich, Germany 
}
\email{frey@lmu.de}

\begin{abstract}
Cytoskeletal networks form complex intracellular structures.  Here we investigate a minimal model for filament-motor mixtures in which motors act as depolymerases and thereby regulate filament length. Combining agent-based simulations and hydrodynamic equations, we show that resource-limited length regulation drives the formation of filament clusters despite the absence of mechanical interactions between filaments.
Even though the orientation of individual remains fixed, collective filament orientation emerges in the clusters, aligned orthogonal to their interfaces.
\end{abstract}

\maketitle

The microtubule cytoskeleton plays an important role in numerous cellular functions such as intracellular transport and cell division~\cite{Nogales2000, Fletcher2010}.
These complex processes are based on active remodelling of the cytoskeletal structure \cite{Howard2009a}, which is mediated by the interaction of microtubules with a variety of microtubule associated proteins (MAPs)~\cite{Howard1997, McAinsh2014, Petry2016b}.
In addition to generating forces between microtubules \cite{Shelley2016}, MAPs play an important role in regulating the length of individual microtubules by affecting the rates of their polymerisation kinetics from tubulin subunits~\cite{Helenius2006, Brouhard2008, Varga2009, Howard}.
How forces affect the large-scale self-organization of microtubules has been studied in detail both theoretically and experimentally~\cite{Leibler1997, Surrey2001, Julicher2000, Sekimoto2004, Shelley2015, Foster2015,Furthauer2019}. 
In contrast, the role of length regulation has only been investigated in the context of individual filaments \cite{Varga2006,Howard2006a,Brouhard2008,Hough2009, Reese2011,Melbinger2012,Kuan_2013,Reese2014, Rank2018}, or of a globally accessible pool of constituents (tubulin and MAPs)~\cite{Doubrovinski2007, Goehring2012a, Ishihara2016b,Zanic2020}.
However, recently the focus of interest is shifting to their role in many filament systems, as there is increasing experimental evidence that this regulatory function, in combination with the local availability of MAPs and tubulin, plays an essential role in the self-organization, scaling and maintenance of microtubule structures~\cite{Good2013b, Hazel853, Reber2013, Gasic, Milunovic2018, Lacroix2018a, Brownlee2019}. 
It remains an important open question how the interplay and spatial redistribution of these resources through cytosolic diffusion and transport along microtubules affects the self-organization of the microtubule cytoskeleton~\cite{Ishihara2020, Geisterfer2020, Ohi2021}. 

\begin{figure}[tbp]
\centering
\includegraphics[width=8.6cm]{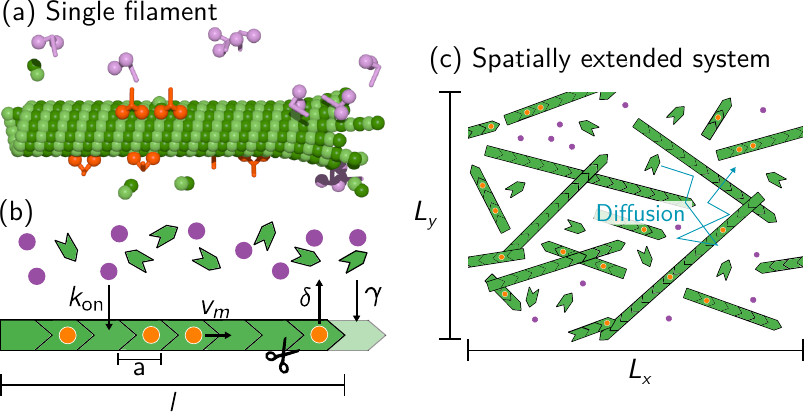}
\caption{%
    Agent-based model.
    (a) Illustration of a filament interacting with a finite amount of tubulin (green) and motor proteins. Motors can be either cytosolic (purple) or filament-bound (orange). 
    (b) Model representation of a single protofilament.
    (c) Illustration of a filament-motor mixture in a box geometry with periodic boundary conditions.
    }
\label{fig:model}
\end{figure}

Here, we approach this question by studying the collective motor-filament dynamics in planar geometry with limited resources of tubulin units and molecular motors. 
These cytosolic resources are spatially redistributed by diffusion while filament-bound motors additionally move uni-directionally towards the filament plus-end where they act as depolymerases (Fig.~\ref{fig:model}).
We show that the interplay of motor-catalyzed depolymerization and local resource availability leads to self-organization of the filament assembly into aster-like patterns. Those patterns show colocalisation of microtubule plus ends and polarity sorting at the interfaces of emerging filament clusters.

\textit{Model.\;---}
We propose an agent-based model that builds on current experimental and theoretical studies addressing the resource-limited length regulation of a single microtubule by the kinesin-8 homologue Kip3 from \emph{Saccharomyces cerevisiae}. \cite{Rank2018}.
Specifically, we study filament dynamics in 
planar geometry
containing a finite number of tubulin units ($N_{\mathrm{T}}$), molecular motors $(N_{\mathrm{M}})$, and filaments $(N_{\mathrm{F}})$; see Fig.~\ref{fig:model}(b).
Each individual filament $i \in {1, \ldots, N_F}$ is represented by a directed rigid rod with fixed minus-end position $\mathbf{b}_{i}$ and fixed orientation $\theta_{i} \in [0,2\pi)$, which are drawn randomly from uniform distributions. 
The lengths $l_{i}(t)$ of the individual filaments are dynamic variables that change by polymerization kinetics at the plus end. 
When filaments shrink to zero length, they are assumed to regrow form the same minus-end position and with the same orientation; filament shrinkage to zero length, though, rarely occurs.
In the cytosol, both motors and tubulin units diffuse freely with diffusion constants $\DM$ and $\DT$, respectively. 
Cytosolic motors can bind with rate $\kon$ to any point that is within the binding radius $r_\mathrm{M}$ along a filament; for details see Supplemental Material Sec.~SII. 
Filament-bound motors move towards the filament plus-end at speed $\vm$, where they catalyse filament depolymerization at rate $\delta$ [see Fig.~\ref{fig:model}(b)]. 
Upon depolymerization, the filament length is reduced by one tubulin unit (of length $a$) and both the plus-end-bound motor and the associated tubulin unit are released into the cytosol.
Cytosolic tubulin within a distance $r_{\mathrm{T}}$ of a filament plus-end, binds to it at the rate $\gamma$, increasing filament length by $a$ [Fig.~\ref{fig:model}(d)].

\textit{Single-filament dynamics.\;---}
Consider a cytosolic volume $V_{0}$, containing a single filament and a finite number of tubulin units
$\rhoT V_{0}$ and motor proteins $\rhoM V_{0}$. 
For now, we assume for simplicity that the cytosolic concentrations $\cM$ and $\cT$ are spatially uniform; this assumption is relaxed when we discuss a spatially extended system with many filaments. 
The length change of the filament is determined by the antagonism between polymerization and depolymerization kinetics $\partial_{t}l (t) = \vg - \vs$ with the growth and shrinkage velocity given by $\vg = a \, \cT \, \gamma$ and $\vs = a \, m^+(t) \, \delta$, respectively, where $m^+(t)$ denotes the density of motors bound to the plus end \cite{Reese2011,Reese2014}.

\begin{figure}[tbp]
\centering
\includegraphics[width=8.6cm]{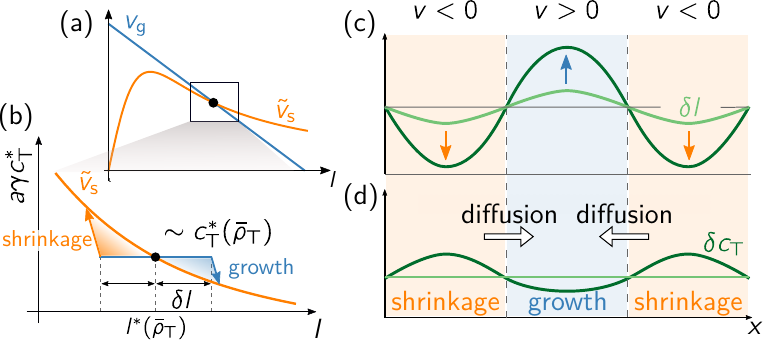}
    \caption{%
    (a) Shrinkage velocity $\vsqss$ and growth velocity $\vg$ as a function of the filament length $l$ with the steady state length $l^*$ determined by the intersection point(s) of $\vg$ and $\vsqss$.  
    (b)-(d) Graphical analysis of the lateral instability. 
    }
\label{fig:instability}
\end{figure}

For biologically relevant parameter ranges, the motor dynamics are fast compared to filament growth and shrinkage \cite{Varga2009,Rank2018}.
This separation of time scales implies that for a given filament length, the motor density can be assumed to be in a quasi-steady state, where the total attachment flux of motors onto the filament, $j_\text{on} = \kon \, \cMqss \, l$, and the off-flux due to depolymerization events at the plus end, $j_\text{off}= \vsqss/a$, are in balance; quasi-steady states are indicated by a tilde.
Thus, the depolymerization velocity $\vsqss = a \, \kon  \, l \, \cMqss$ is determined by the cytosolic density $\cMqss$, which in turn is related to the filament-bound motor number $\tilde{M}$ via mass conservation $\rhoM V_{0} = \cMqss V_0 + \tilde{M}$. In steady state, the filament-bound motor density exhibits
an antenna profile $\tilde{m}(s) = \frac{\kon \cMqss}{\vm} \,s$ \footnote{Note the occupation density of motors at the filament plus-end is not equivalent to $\tilde{m}(l)$. The filament-bound motor density exhibits a boundary layer such that $\vm\tilde{m}(l)=a m^{+}\delta$ \cite{Parmeggiani2003},\cite{Parmeggiani2004}}, which is inferred from the transport equation $\partial_t m(s,t) = -\vm \partial_s m(s,t) + \kon \cM(t)$ \cite{Parmeggiani2003, Parmeggiani2004}, implying $\tilde{M} = \frac{\kon \cMqss}{2\vm} \, l^2$. 
Taken together, we find a relation for the shrinkage velocity in terms of the filament length  and the total motor concentration 
\begin{equation}
    \vsqss (l, \rhoM) 
    = a \, \kon \, l \, \cMqss 
    = a \, \kon \, l \, \frac{\rhoM}{1 + l^2/l_c^2} 
    \, ,
\label{eq:vshrinkage_steady}
\end{equation}
where we have defined the characteristic length scale $l_\mathrm{c} := \sqrt{2 \vm V_{0}/\kon}$.
For filament lengths $l < l_\mathrm{c}$, the shrinkage velocity increases with the filament length, as would be expected with unlimited motor resources and has been observed experimentally \cite{Varga2006,Varga2009}.
Increasing the filament length beyond $l_{\mathrm{c}}$ leads to a depletion of the cytosolic motor pool and thereby a decreased on-flux $\kon \, \cMqss \, l$.
According to the flux balance condition, this reduces the off-flux $\vsqss/a$ and thus the shrinkage velocity, so that $\vsqss \sim 1/l$ for $l \gg l_{\mathrm{c}}$ [see Fig.~\ref{fig:instability}(a)].

The growth velocity $\vg$ can be written in terms of filament length $l$ and total tubulin density $\rhoT$ using tubulin mass conservation 
($\rhoT V_0 = \cT V_0 + l/a$) as $\vg(l,\rhoT) = \gamma \, ( \rhoT \, a - l /V_0)$.
Together with the balance of filament  growth and shrinkage, $\vg(l,\rhoT) = \vsqss(l,\rhoM)$, this determines the steady state length $l^*(\rhoT, \rhoM)$ 
[Fig.~\ref{fig:instability}(a)] \footnote{Depending on the functional form of the shrinkage velocity $\vsqss(l)$ the dynamics is either monostable with a single steady state length or bistable; see Supplemental Material Sec.~SIV for details.}.

\textit{self-organization in a spatially extended system.\;---}
How does the length regulation of individual filaments play out in a spatially extended system where resources are shared by cytosolic diffusion between many filaments?
In the limiting case where the cytosolic concentration is slowly varying on the scale of the (typical) filament length, the filaments can be treated as point-like objects carrying a tubulin mass proportional to their length $l(\mathbf{x},t)$. 
The single filament dynamics can then immediately be generalized to a local length regulation dynamics
\begin{equation}\label{eq:dynamics_l}
    \partial_{t}l(\mathbf{x},t) 
    =  a \, \gamma \, \cT(\mathbf{x},t) 
    - \vsqss(\mathbf{x},t) \,,
\end{equation}
with the local shrinkage speed given in terms of the local quasi-steady state approximation for the cytosolic motor density, $\vsqss(\mathbf{x},t) = a \,  \kon \, l(\mathbf{x},t)\, \cMqss[l(\mathbf{x},t),\rhoM(\mathbf{x},t)]$ (cf. Eq.~\ref{eq:vshrinkage_steady}). The dynamics of the cytosolic tubulin concentration is governed by a reaction-diffusion equation
\begin{equation}\label{eq:dynamics_cT}
    \partial_{t} \cT(\mathbf{x},t) 
    = 
    \DT \nabla^{2}\cT(\mathbf{x},t) - \frac{\gamma \cT(\mathbf{x},t) - \vsqss(\mathbf{x},t)/a}{V_0} \, , 
\end{equation}
where the local polymerization kinetics induces sinks and sources of cytosolic tubulin; here $V_0 = V/N_\mathrm{F}$ denotes the cytosolic volume associated with a single filament. 
The total motor density is redistributed by cytosolic diffusion
\begin{align}
\label{eq:dynamics_rhoM}
    \partial_t \rhoM (\mathbf{x},t)
    =
    \DM \nabla^2 \cMqss  [l(\mathbf{x},t),\rhoM(\mathbf{x},t)] 
    \, , 
\end{align}
where we again used the local quasi-steady state approximation for the cytosolic motor density $\cMqss (\mathbf{x},t)$.
Taken together, Eqs.~\eqref{eq:dynamics_l}--\eqref{eq:dynamics_rhoM} form a closed set governing the system's dynamics in the long-wavelength limit.

The stability of a spatially uniform state $(l^*,\cT^*,\barRhoM)$ against spatial perturbations can be studied using  
a linear stability analysis (see Supplemental Material Sec.~SIV for details).
Figure~\ref{fig:linear-stability} shows a typical dispersion relation $\sigma(q)$ for the eigenvalue with the largest real part and the ensuing stability diagram as a function of $\barRhoM$ and $\barRhoT$. 
For $\barRhoT > \barRhoT^\mathrm{crit}(\barRhoM)$ there is a band of unstable Fourier modes $q \in (0,q_\mathrm{max})$ extending to long wavelengths ($q \to 0$).
It is instructive to first consider the particular limit of well-mixed cytosolic tubulin.
Then, the marginal mode $q_\mathrm{max}$ reduces to $q_\mathrm{max}^2 = -\rhoM \partial_l \vsqss|_{l^*}/(\DM \cMqss)$. 
This implies that there is an instability against spatial perturbations (lateral instability) only if $\partial_l \vsqss|_{l^*} < 0$. 
Moreover, the band of unstable modes narrows with increasing $\DM$, showing that cytosolic motor diffusion attenuates the lateral instability. 
Relaxing the assumption of well-mixed cytosolic tubulin, i.e., explicitly accounting for tubulin diffusion, yields the critical ratio of diffusion constants $\DT^\mathrm{crit}/\DM \approx \gamma / (a \kon V_0 \barRhoM)$ in the limit $l^* \gg l_\mathrm{c}$.
For physiological parameters, we find that the there is a lateral instability if the average number of motors per filament satisfies $\barRhoM^\mathrm{crit} V_0 > 0.57 \, \DM/\DT$. 
This condition is well met for biologically relevant motor concentrations as $\DM/\DT \sim 1/6$ (see Supplemental Material Sec.~SI).

\begin{figure}[tbp]
\centering
\includegraphics[width=8.6cm]{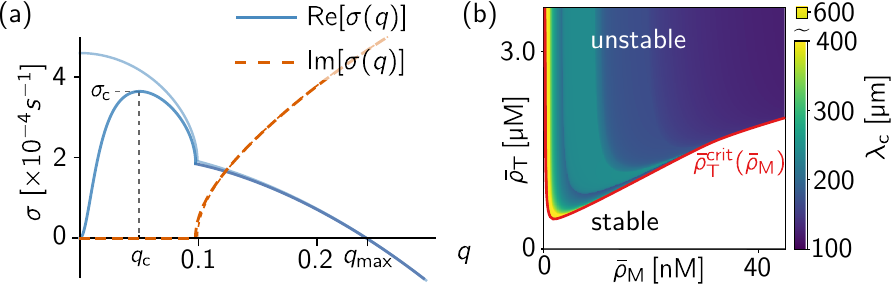}
    \caption{%
    (a) Leading eigenvalue $\sigma(q)$ for $\barRhoM = \SI{50}{\nano M}$, $\barRhoT = \SI{2.75}{\micro M}$ , $\DM = \SI{0.5}{\micro \m^2\s^{-1}}$ and $\DT = \SI{6}{\micro \m^2\s^{-1}}$; other parameters are specified in the Supplemental Material Sec.~SI. The dispersion relation in the limit of well-mixed cytosolic tubulin is shown in light blue.  
    (b) Stability diagram and wavelength of the fastest growing mode $q_\mathrm{c}$ in the $(\barRhoM,\barRhoT)$-parameter space. The boundary of the laterally stable parameter regime, $\barRhoT^\mathrm{crit}(\barRhoM)$, is shown in red. 
    }
\label{fig:linear-stability}
\end{figure}

The feedback mechanism underlying the lateral instability can be explained in terms of a mass-redistribution instability \cite{Halatek2018,Halatek2018a,Brauns2020}.
To simplify the argument, we  set $\DM = 0$ for the moment so that the total motor density remains invariant under the dynamics and therefore  spatially uniform $\rhoM = \barRhoM$, cf. Eq.~\eqref{eq:dynamics_rhoM}. 
Consider now a small perturbation $\delta l(\mathbf{x})$ added to the homogeneous state $l^*$,  while keeping the cytosolic tubulin concentration $\cT(\mathbf{x}) =\cT^*$ initially constant [Fig.~\ref{fig:instability}(b,c)].
Since then $\vg = a\, \gamma \, \cT$ initially remains uniform, the effect of $\delta l(\mathbf{x})$ on the net growth velocity  $v = \vg - \vsqss$ depends on the slope of the shrinkage velocity at $l^*$. 
For $\partial_l \vsqss|_{l^*} < 0$, 
filaments grow (shrink) when they are long (short).
This leads to an  decrease (increase) of the cytosolic tubulin concentration [arrows in 
Fig.~\ref{fig:instability}(b,c)]
creating gradients in the cytosolic tubulin concentration that drive diffusive transport of tubulin mass towards regions of increased filament length. 
Since this tubulin mass redistribution leads to an increase of $\vg$ in regions where $\delta l > 0$, it promotes further filament growth there, i.e., the initial spatial perturbation $\delta l(\mathbf{x})$ is amplified  [Fig.~\ref{fig:instability}(b)].
In contrast, if the regulatory kinetics is such that the shrinkage velocity increases with filament length ($\partial_{l}\vsqss |_{l^*} > 0$), the effect is opposite. 
Cytosolic tubulin diffusion then redistributes the tubulin mass to regions with shorter filaments, counteracting the original disruption.
Taken together, one finds the condition $\partial_{l}\vsqss|_{l^*} < 0$ for a spatial instability that is driven by free tubulin diffusion, in accordance with the result of the linear stability analysis. 

The above reasoning also explains why cytosolic diffusion of motor proteins mitigates the lateral instability. 
Regions with short filaments contain fewer binding sites for motors and thus the cytosolic motor concentration is high there. 
The opposite holds for regions with long filaments. 
This creates gradients, and thereby diffusive fluxes, of motors towards regions of long filaments. 
The resulting diffusive influx of motors increases the rate of filament depolymerization there and thus counteracts the instability driven by tubulin diffusion.

\begin{figure*}[tbh]
\includegraphics{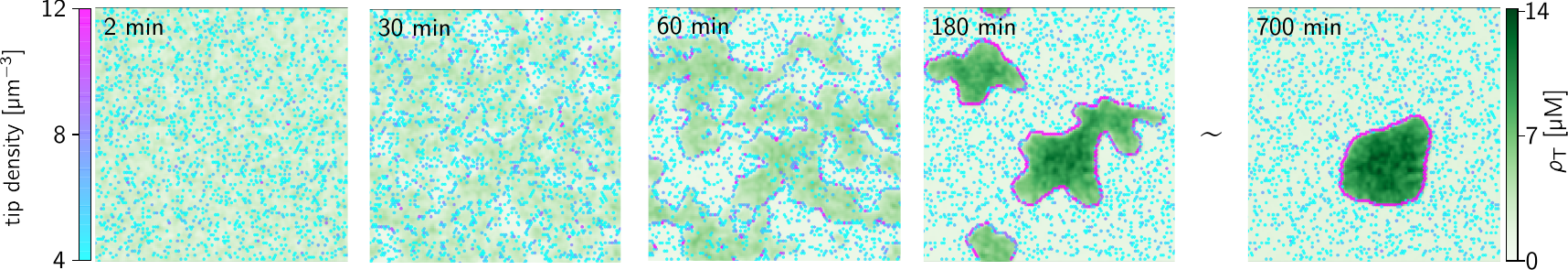}
    \caption{%
    Snapshots of the total tubulin density $\rhoT(\mathbf{x},t)$  and the tip density.
    Parameters are as in Fig.~\ref{fig:linear-stability}(a); $L_\mathrm{x} = L_\mathrm{y} = \SI{150}{\micro \m}$.
    }
    \label{fig:snapshots}
\end{figure*}

\textit{Agent-based simulations.\;---}
To study the spatio-temporal dynamics above the critical tubulin concentration $\barRhoT^\mathrm{crit}(\barRhoM)$ we perform agent-based simulations.  
While Eqs.~\eqref{eq:dynamics_l}--\eqref{eq:dynamics_rhoM} capture well the initial dynamics at the long-wavelength instability, they fail to give the correct dynamics once gradients begin to emerge at the small length scales (see  Supplemental  Material  Sec.~SIV).
What this continuum theory lacks are effects due the spatial extent of the filaments which includes motor binding along the length of filaments as well as motion of each filament plus-end due to polymerization kinetics.

Figure~\ref{fig:snapshots} shows a time sequence obtained from the simulations (see also Movie S1).
First, regions with short (depletion zones) and long (clusters) filaments are formed, which corresponds to the initial dynamics described by the mass redistribution instability (Fig.~\ref{fig:snapshots}, $t=\SI{30}{\min}$). 
Moreover, filament plus-ends start to accumulate at the interface between these zones.  
As the dynamics progresses, the depletion zones grow in size and the interfaces sharpen (Fig.~\ref{fig:snapshots}, $t=\SI{60}{\min}$).
At this time point, the filament-length distributions match on a qualitative level with experimental measurements \cite{Rank2018} (see Supplemental Material Sec.~SIII). 
Subsequently, the high density regions segregate into individual large scale filament clusters, which are characterized by sharp boundaries and strong co-localization of filament plus-ends at  their periphery  (Fig.~\ref{fig:snapshots}, $t = \SI{180}{\min}$). 
This co-localization is caused by the movement of the filaments' plus-end due to polymerization dynamics that is directed to zones where the net growth rate changes sign, namely cluster interfaces.
In the long run, the large filament clusters grow at the expense of the smaller ones, until eventually only a single cluster remains, which then develops into an aster-like structure 
(Fig.~\ref{fig:snapshots}, $t = \SI{700}{\min}$).

\begin{figure}[!b]
\centering
\includegraphics[width=8.6cm]{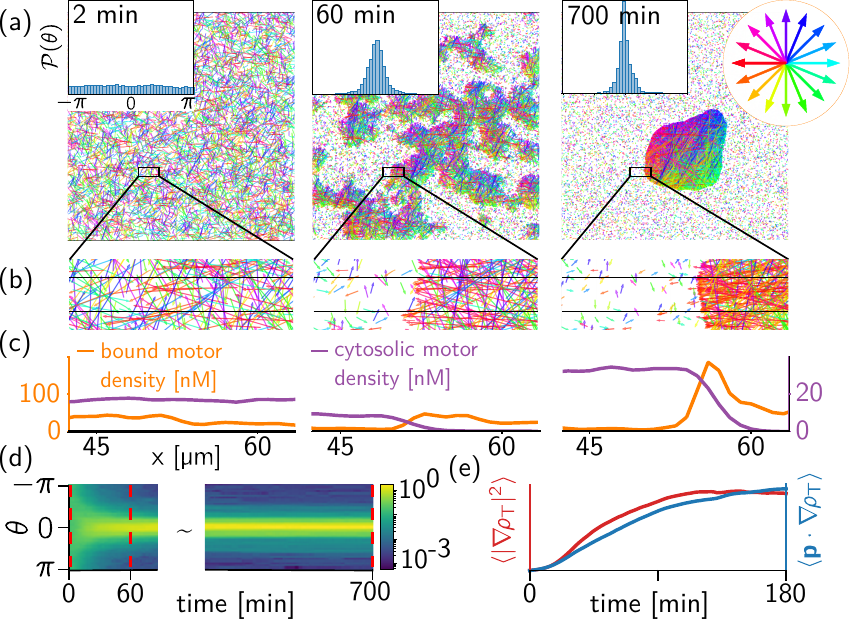}
    \caption{%
    (a) Snapshots of filament arrangement in Fig.~\ref{fig:snapshots}. Filaments are color coded according to their orientation (color wheel); insets show the weighted distribution $\mathcal{P}(\theta)$.
    (b) Zoom into a structure interface. 
    (c) Filament-bound (orange) and cytosolic (purple) motor concentration averaged along the vertical direction for the area enclosed by the black windows in (b).
    (d) Kymograph of $\mathcal{P}(\theta)$ with the (logarithmic) color scale showing the normalized (by area) frequency of the measured angles; the dashed, red lines correspond to the insets in (a).
    (e) Time trace of $\langle\mathbf{p} \cdot \nabla \! \rhoT \rangle$ and $\langle|\nabla \! \rhoT|^2\rangle$ (ordinate in a.u.).
    }
\label{fig:polarity-sorting}
\end{figure}

Inside the clusters, the filaments exhibit net polar order that is aligned along tubulin-density gradients, i.e., orthogonal to the cluster boundaries.
This is because the plus ends localized there belong predominantly to filaments whose minus end lies within the cluster's interior, implying an orientation orthogonal to the boundary on average (Fig.~\ref{fig:polarity-sorting} and Movie S2).

To quantify this effect, we monitor the density gradient $\nabla \! \rhoT$, the local net polarity $\mathbf{p}$, and the angle $\theta$ enclosed between these vectors.
Figure~\ref{fig:polarity-sorting}(d) shows the time evolution of the histogram $\mathcal{P}(\theta)$ of the angle $\theta$ weighted by the product of the magnitudes of $\nabla \! \rhoT$ and $\mathbf{p}$ to highlight the alignment of filaments near the cluster boundaries.
The initially uniform distribution $\mathcal{P}(\theta)$ evolves quickly into a peaked distribution centered around zero---indicating the onset of polar order---and subsequently sharpens slowly; see also snapshots in insets of Fig.~\ref{fig:polarity-sorting}(a). 
The onset of this polar order occurs simultaneously with the mass-redistribution instability, as can be seen from the comparison of the spatial averages $\langle \mathbf{p} \cdot \nabla \! \rhoT \rangle$ and $\langle |\nabla \! \rhoT|^2 \rangle$, which are coarse-grained measures of filament orientation and density gradients, respectively; Fig.~\ref{fig:polarity-sorting}(e). 

The polar order leads to advective flow of filament-bound motors out of clusters, which is balanced against diffusive influx caused by gradients in cytosolic motor concentration [see Fig.~\ref{fig:polarity-sorting}(c)].
Fast binding of motors inside clusters together with advective motor transport leads to the depletion of motors in the cluster interior and the formation of sharp gradients in cytosolic motor concentration.
Those gradients help to maintain the filament plus-end localization at the interfaces: Plus ends that protrude beyond the interface are subjected to an increased on-flux of motors, causing the filaments to shrink back. 
Conversely, plus-ends within the cluster are subjected to a reduced motor on-flux, causing them to grow towards the interface.
Finally, the sharp cytosolic gradient leads to a shrinkage velocity $\vsqss$ that is independent of filament length because motor attachment occurs only in a narrow band at the interface.
This is a collective effect and in contrast to the length regulation of a single filament, which is strongly length-dependent [cf.\ Eq.~\eqref{eq:vshrinkage_steady}].
The size regulation of clusters and a quantitative analysis of the final, aster-like, stationary state will be presented in a forthcoming publication \cite{StriebelXXX}.

\textit{Discussion.\,---}
Commonly, the spatial self-organization of a filament arrangement is attributed to motor proteins that reorient and move filaments by mechanical forces, such as dynein or kinesin-5 \cite{Shelley2016,  TAN2018, Furthauer2019, Striebel2020}.
Here, we have shown that microtubule length regulation (through kinesin-8) in combination with resource limitation can lead to aster-like spatial patterns.
The underlying instability is driven by diffusive redistribution of cytosolic tubulin mass.
Such a mass-redistribution instability is a potential candidate to explain the emergent self-organization observed in cell extracts \cite{Cheng2019}. Notably, this self-organization is heralded by spatial patterns that emerge in the tubulin density, which have comparable morphology and wavelength ($\sim \SI{100}{\micro \m}$) as those we observe in our simulations (cf.\ Fig.~\ref{fig:snapshots}).
We also expect that our theory for resource-limited filament length regulation can be used to investigate heterogeneous growth dynamics in systems where spatial heterogeneities in filament length and/or density are imposed, e.g.\ by experimental design \cite{Geisterfer2020} or by upstream gradients \cite{Oh2016,Decker2017a}.
In the system studied here, inhomogeneous filament polymerization occurs spontaneously, leading to inhomogeneous filament densities (lengths).

From a broader perspective, the conceptual model investigated here is in itself an interesting active matter system exhibiting self-organized patterns, polarity sorting, and coarsening. Combining (active) length regulation with mechanical interactions of filaments will be an important avenue for future research.

\begin{acknowledgments}
We would like to thank Henrik Weyer, Isabella Graf and Philipp Geiger for careful reading of the manuscript.
We acknowledge financial support by the Deutsche Foschungsgemeinschaft through the Excellence Cluster ORIGINS under Germany's Excellence Strategy (EXC-2094-390783311). 
\end{acknowledgments}

\clearpage \newpage

\begin{widetext}
\newgeometry{
 	left = 35mm,
  	right = 35mm,
  	top=35mm, 
  	bottom=45mm, 
  	headsep=30pt
}
\setstretch{1.1}

\makeatletter
\c@secnumdepth=3
\makeatother

\renewcommand{\thesection}{S\Roman{section}} 
\renewcommand{\thefigure}{S\arabic{figure}} 

{\centering\LARGE\scshape Supplemental Material\par}

\section{Parameter estimation}

Since experimental data are available for many of the parameters in our model \cite{Varga2009,Hyman1995,Brouhard2008,McIntosh1984}, we can assign experimentally motivated numerical values to them. 
In the following we will show how to convert the experimentally measured rates into appropriate rates for our simulation.
The experimental and the resulting model parameters are summarized in table~\ref{tab:parameters}. 

A typical microtubule consists of thirteen protofilaments.
We simplify our analysis by considering only a single protofilament per microtubule, i.e., we neglect correlations between neighbouring protofilaments of a single microtubule. 
This is achieved by converting rates and resources to rates and resources per protofilament.
Given a height  $\tilde{L}_z$ of our simulation environment,
the total number of resources per protofilament is given by
$
N_\mathrm{M} = \bar{\rho}_\mathrm{M} L_x L_y \tilde{L}_z/13 \; \text{ and }
N_\mathrm{T} = \bar{\rho}_\mathrm{T} L_x L_y \tilde{L}_z/13 \;
$
for motor proteins and tubulin units respectively.
We define $L_z := \tilde{L}_z/13$ as the effective $z$-extent per protofilament, which corresponds to the effective height of our simulation box. 
For a given choice of the spatial extents of the simulation box the values for $N_\mathrm{T}$ and $N_\mathrm{M}$ will be determined from the experimentally given values for the densities $\barRhoT$ and $\barRhoM$. 

The \textit{motor attachment rate} $\kon$ can be determined from the experimentally measured rate constant for motor attachment, which is given in units per concentration, per time, and per length.
Varga et al.\ \cite{Varga2009} report a value of $\kon^\mathrm{exp} = \SI{0.4}{\nano M^{-1} s^{-1} \micro \m^{-1}} = \SI{0.66}{\micro \m^3 s^{-1} \micro \m^{-1}}$, which was measured over a range of motor concentrations; note that we do not cancel one unit of \si{\micro \m} to emphasize that $\kon$ is a rate per volume concentration and per length of the microtubule. The motor on-flux onto a filament of length $l$ at a given motor concentration $\cM$ is than given by $\kon^\mathrm{exp}\cM l$. The motor concentration in experiments is measured in \si{\nano M} which is converted into a number densities as $\SI{1}{\nano M} \approx \SI{0.6}{\mathrm{particles}/{\micro \m^3}}$. 
Since Varga et al.\ used a TIRF setup, where motors can bind to and walk on approximately 5 protofilaments \cite{SCHNEIDER2015}, the attachment rate per protofilament is given by $\kon = \kon^\mathrm{exp}/5 \approx \SI{0.13}{\micro \m^3 s^{-1} \micro \m^{-1}}$. 
For our agent-based simulations of the filament-motor mixture, we need to further convert this rate to a per-capita rate for a single motor protein. 
To this end, we first convert the experimental rate constant per volume concentration into a rate per area concentration by $\kon^{2\mathrm{D}} = \kon^\mathrm{exp}/L_z$.
This is converted to a per-capita rate by specifying a reaction radius $r_\mathrm{M}$ within which the motors can attach to the filament.  
Thus, the experimental value for the attachment rate constant per protofilament $\kon^\mathrm{exp}$ can be converted to a per-capita rate used in the simulations by $\kon^\mathrm{sim}=\kon^\mathrm{exp}/(L_z \pi r_\mathrm{M}^2)$. 

Given the experimental value for the attachment rate, this scaling relation gives us some freedom in choosing the effective height $L_\mathrm{z}$ of the simulation box and the reaction radius $r_\mathrm{M}$. The height of the simulation box has to be chosen such that all concentrations (tubulin, motor and filament concentration) can be assumed to be well mixed in $z-$direction, i.e., $L_z \ll \lambda_\mathrm{c}$, where $\lambda_\mathrm{c}$ is the wavelength of the initial instability (see Sec.~\ref{sec:LSA}). 
Moreover $L_\mathrm{z}$ has to be chosen large enough such that the number of filaments and particles is high enough to prevent stochastic effects due to number fluctuations.
As long as these constrains are fulfilled we are free to choose $L_z$ in a way convenient for our simulations. In particular, this means we choose $L_z$ small to keep the particle numbers sufficiently low for the numerical simulations to be feasible.
The reaction radius $r_\mathrm{M}$ must be smaller than the average distance covered by a motor in the time interval $\Delta t$ by diffusion, where $\Delta t$ is the time increment of the simulation (see Sec.~\ref{sec:simulation}). Otherwise particles could cover an unphysically large distance in the time interval $\Delta t$ by successive attachment/ detachment events.

The \textit{spontaneous polymerization rate} $\gamma$ of microtubules can be obtained similarly to the attachment rate. 
The polymerization velocity of microtubules has been measured to be $v^\mathrm{exp} = \SI{0.19}{\micro \m \min^{-1} \micro M^{-1}}$ \cite{Brouhard2008}, which corresponds to a polymerization rate constant $\gamma^\mathrm{exp} = v^\mathrm{exp}/a = \SI{0.38}{\micro M^{-1} s^{-1}} = \SI{6.3e-4}{\micro \m^{3}\s^{-1}}$. Analogous to the case of the attachment rate constant, the polymerization rate constant depends on the volume concentration of tubulin units. We convert this into a per-capita polymerization rate by $\gamma^\mathrm{sim} = \gamma^\mathrm{exp}/(L_z \pi r_\mathrm{T}^2)$, where $r_\mathrm{T}$ denotes the reaction radius of a tubulin unit.

For the simulation results shown in the main text, we choose $r_\mathrm{T} = r_\mathrm{M} = \SI{0.04}{\micro \m}$ and $L_z = \SI{0.6}{\micro \m}$. At  average concentrations $\barRhoT = \SI{2.75}{\micro M}$ and $\barRhoM = \SI{50}{\nano M}$ this results in a total number of $N_\mathrm{T} \approx \SI{2.2e7}{}$ tubulin units and $N_\mathrm{M} \approx \SI{4e5}{}$ motor proteins; as well as per-capita binding rates $\gamma^\mathrm{sim} \approx \SI{0.22}{\s^{-1}}$ and $\kon^\mathrm{sim} \approx \SI{40}{\s^{-1}\micro \m^{-1}}$ (see Table~\ref{tab:parameters}).

\setlength{\tabcolsep}{5pt}
\begin{table}[!t]
\centering
\begin{tabular}{@{} *6l @{}} \toprule
 &  & Experiment & Theory & Simulation & Ref. \\
\midrule
\multicolumn{3}{@{}l}{Motor parameters} & \\
\midrule
    Motor velocity &$\vm$  &$ \SI{0.053}{\micro \m\,\s^{-1}}$ & $\SI{0.06}{\micro\m\,\s^{-1}}$ & $\SI{0.06}{\micro\m\,\s^{-1}}$ &\cite{Varga2009}  \\
    Cytosolic motor diffusion  &$\DM$  & --- & $\SI{0.5}{\micro \m^2 \s^{-1}}$  & $\SI{0.5}{\micro \m^2 \s^{-1}}$ & --- \\ 
    Attachment rate &$\kon$  &$ \SI{0.13}{\micro \m^3 \s^{-1}  \micro\m^{-1}}$ & $\SI{0.12}{\micro \m^3 \s^{-1}  \micro\m^{-1}}$  &$\SI{40}{\s^{-1}  \micro\m^{-1}}$& \cite{Varga2009}\\
    Detachment rate & $\koff$  &$\SI{5e-3}{\s^{-1}}$ & $\SI{0}{s^{-1}}$& $\SI{0}{s^{-1}}$  & \cite{Varga2009}\\
    Depolymerization rate & $\delta$  &$\SI{2.3}{\s^{-1}}$ &--- ($\SI{2.3}{\s^{-1}}$) &--- ($\SI{2.3}{\s^{-1}}$)& \cite{Varga2009}\\
\midrule
\multicolumn{3}{@{}l}{Tubulin and filament parameters} & \\
\midrule 
     Size of a tubulin dimer & $a$ & $\SI{8.4}{\nano \m}$ & $\SI{8.4}{\nano \m}$  &$\SI{8.4}{\nano \m}$& \cite{Hyman1995} \\ 
     Polymerization rate & $\gamma$ &  $\SI{6.3e-4}{\micro\m^3\s^{-1}}$& $\SI{1e-3}{\micro \m^3\, \s^{-1}}$&$\SI{0.33}{s^{-1}}$& \cite{Brouhard2008} \\
     Cytosolic tubulin diffusion & $\DT$ & $\SI{6}{\micro \m^2 \s^{-1}}$ & $\SI{6}{\micro \m^2 \s^{-1}}$ & $\SI{6}{\micro \m^2 \s^{-1}}$&\cite{McIntosh1984} \\
     Volume per filament & $V_0$ & --- & \SI{0.5}{\micro\m^{3}} & \SI{0.5}{\micro\m^{3}}&---  \\
\bottomrule
\end{tabular}
\caption{Experimentally measured parameters for the kinesin-8 homolog Kip3 from \emph{Saccharomyces cerevisiae} and parameters chosen in our theoretical discussion. 
The experimental parameters are meant to provide the correct order of magnitude. 
The qualitative results of our analysis are not sensitive to the choice of specific parameters (as long as the parameters are in a comparable range). To convert experimental rates into rates required for our simulation we used $r_\mathrm{M} = r_\mathrm{T} = \SI{0.04}{\micro \m}$ and $L_z = \SI{0.6}{\micro \m}.$ 
If not explicitly stated otherwise we used parameters as specified in the column Theory.}
\label{tab:parameters}
\end{table}

The diffusion constant of cytosolic tubulin was measured as $\DT = \SI{6}{\micro m^2.s^{-1}}$ \cite{McIntosh1984}. 
To the best of our knowledge, the cytosolic diffusion constant $\DM$ of Kip3 is not known.
Since the molecular mass of Kinesin-8 is about $2$--$3$ times larger than that of tubulin \cite{CellBiology}, we expect the diffusion constant of Kip3 to be of the order $\DM \sim \SI{1}{\micro m^2.s^{-1}}$.
As we discuss in Sec.~\ref{sec:LSA} below, the precise value of the motor diffusion is not relevant for the qualitative results of our model. Linear stability analysis predicts that slower motor diffusivities entail a shorter wavelength of the fastest growing mode (see Fig.~\ref{fig:DM-DT-stability-diagram}).
To keep the computational cost of our agent-based simulations reasonable, we choose $\DM = \SI{0.5}{\micro m^2.s^{-1}}$, which allows us to simulate a smaller system and thus keep the number of particles in the simulation volume manageable.

\section{Agent-based simulation}\label{sec:simulation}

\subsection{Single-filament simulation}

To simulate the motor-mediated length regulation of a single filament, we follow a two-pronged approach: 
First, we employ the Gillespie algorithm \cite{Gillespie1977} to simulate the full stochastic dynamics of motor proteins using a lattice gas model (TASEP), which has been shown to be a good model system for studying the motion of motor proteins on microtubules.  \cite{Lipowsky2001,Lipowsky2003,Parmeggiani2003,Chou_2011}. 
Since this exact simulation approach suffers from performance problems when studying many filaments coupled via a diffusive reservoir, we also implement an approximate simulation scheme in which filament-bound motors move deterministically and steric interactions between motor proteins are neglected.
To guarantee its validity, the approximate simulation scheme is compared with the results of the exact Gillespie method in the relevant parameter range.

We develop our approximate simulation schema based on theoretic results of the TASEP-LK model \cite{Parmeggiani2003,Parmeggiani2004}: In the limiting case of low densities, the dynamics of the mean motor proteins on the filament is described by the mean-field equation
\begin{equation}
    \partial_{t}m(s,t) = -\vm \partial_{s}m(s,t) +\kon m(s,t) \;. 
\end{equation}
While this result follows strictly from full stochastic dynamics, it can also be obtained from deterministic dynamics where during the time interval $\Delta t$ the position $s_i$ of the motor proteins is updated as 
\begin{equation}
   s_{i}^{n+1} \rightarrow s_{i}^{n} + \vm \Delta t \;, 
\end{equation}
i.e. all motor proteins move ballistically with the same speed $v_m$.
Moreover, for sufficiently high depolymerization rates and low densities, TASEP-LK predicts that the depolymerization rate $\vsqss$ is given by \cite{Rank2018}
\begin{equation}
       \vsqss = \vm \,m(l) [1-m(l)] = a\, m^+\delta \;. 
\end{equation}
Neglecting effects from particle exclusion, this reduces to $\vsqss = \vm \,m(l) = a\, m^+\delta \;$, i.e., instead of explicitly modeling each depolymerization event, one can simply move the motor forward and if $s_{i} > l$ depolymerize the filament.
Both approximations have the key advantage that they can be performed in parallel in the case of a spatially extended system with many motor proteins. 
To ensure that several motors do not depolymerize the filament within one iteration step, the time step must be chosen sufficiently small. 
Throughout all of our simulations we choose $\Delta t = \SI{0.01}{s}$. 
With a motor speed of $\vm = \SI{0.06}{\micro \m \s^{-1}}$, this means that `double' depolymerization events will only occur if the distance between two consecutive motors is less then $\SI{6e-3}{\micro \m}$, which is less then the length $a$ of a tubulin unit. Figure~\ref{fig:approx-vs-gillespie} shows a comparison between the filament length at steady state obtained by a full stochastic simulation and our approximate simulation scheme.  

\begin{figure}[tb]
\centering
\includegraphics{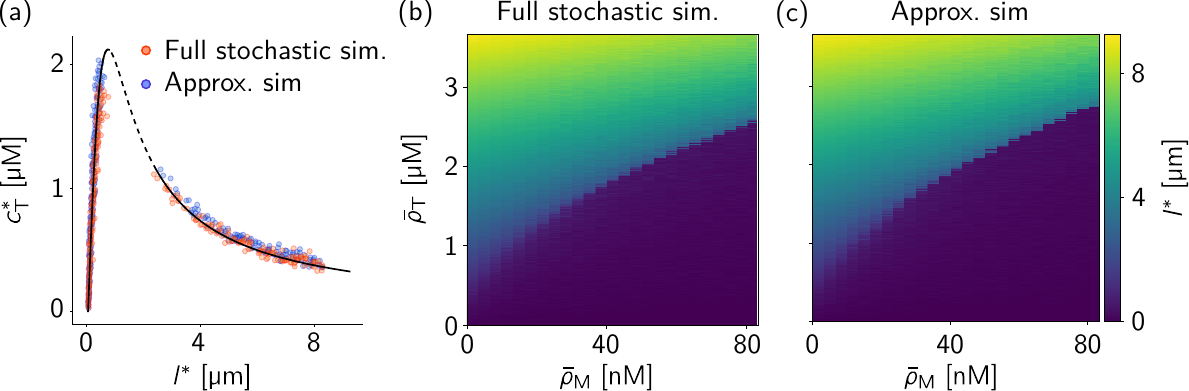}
    \caption{%
Comparison between an exact (Gillespe algorithm) stochastic simulation and our approximate simulation scheme for a single filament. 
For the relevant parameter range, we find good agreement between the approximate and exact simulation schemes.
Filaments were initialized fully polymerized (all tubulin incorporated in the microtubule) for all simulations. Parameters are chosen as specified in table~\ref{tab:parameters}. 
(a) Steady-state cytosolic tubulin concentration as a function of the steady-state filament length for $\rhoM = \SI{30}{\nano M}$.
The symbols indicate simulation results, the solid black line represents the analytical result; see Eq. (1) in the main text. The dashed line indicates the range of unstable fixed points in the bi-stable parameter range (see Sec.~\ref{sec:LSA} and Fig.~\ref{fig:LSA-rhoM-rhoT-diagrams} (b) for an explanation). 
(b)-(c) Filament length $l^*$ in steady state as a function of $\rhoM$ and $\rhoT$. 
The results of the exact and approximate simulation schemes agree very well over the entire parameter range. 
}
\label{fig:approx-vs-gillespie}
\end{figure}      

\subsection{Spatially extended system}

\subsubsection{Filament dynamics}
A filament in the spatially extended system is characterized by its minus end position $\mathbf{b}_{i}$, orientation $\theta_i$, and length $l_i$. 
The positions and orientations of the minus ends are initially drawn from a uniform distribution and remain unchanged throughout the simulation. 
Free tubulin units are inizialized at random positions within the simulation box.
The stochastic dynamics of the free tubulin units is implemented as follows:
First, we check whether filament plus ends are within the distance $r_\mathrm{T}$ of the tubulin unit.  
For each plus end $(0 \dots k)$ within range, a reaction time $\tau_k$ is drawn from an exponential distribution; $\tau_k \sim \gamma^\mathrm{sim} \exp(\gamma^\mathrm{sim} t)$. 
We choose the smallest reaction time $\tau_\mathrm{k}$ that fulfills $\tau_k < \Delta t$, where $\Delta t$ denotes the time increment of the simulation, and perform the respective growth event. 
If there is no filament plus end in range or if all reaction times are $\tau_k > \Delta t$, we let the tubulin unit perform a free diffusive motion implemented by a Brownian dynamics algorithm \cite{grassia_1995}. 
This is the position $\mathbf{x}_i^n = (x_i^n,y_i^n)$ of the tubulin unit is updated as 
\begin{align}
x^{n+1}_i &\rightarrow x^{n}_i + A r^{n}_i \; ,
\\
y^{n+1}_i &\rightarrow x^{n}_i + A r^{n}_i \;.
\end{align}
Here $r^{n}$ are independent and identically distributed random variables with zero mean and $A$ is an amplitude chosen such that the fluctuation dissipation theorem is satisfied \cite{StatisticalMechanicsSchwabl}. 
Importantly, however, it is not necessary for the random numbers to be Gaussian-distributed \cite{grassia_1995}. 
Here, we choose uniform random numbers in the interval $(-0.5,0.5)$ as this has several implementation-specific advantages. 
Since the variance of uniformly distributed random variables is given by $\langle (r^{n})^2\rangle = 1/12$ this results in $A = \sqrt{24 D_T \Delta t}$.
The pseudocode for the stochastic dynamics of cytosolic tubulin units is given in  Algorithm~\ref{alg:fil-growth}.

\subsubsection{Motor dynamics}
The motors can be either filament-bound or free (cytosolic). 
We perform simulations of the stochastic dynamics of cytosolic motors analogous to that of cytocolic tubulin. 
First, we determine all potential binding partners, i.e., all filaments $(0\dots k)$ that intersect with a radius $r_\mathrm{M}$ around the motor position $\mathbf{x}_i = (x_i,y_i)$ (see Fig.~\ref{fig:sim-sketch}). 
Next, the chord length $\Delta l$ is calculated; for an illustration see Fig.~\ref{fig:sim-sketch}(b). 
The per-capita attachment rate for a single motor protein attaching to the filament is then given by $\kon^\mathrm{sim} \Delta l$. 
Similar as for the free tubulin dynamics, the reaction times $\tau_k \sim \kon^\mathrm{sim} \Delta l \exp(-\kon^\mathrm{sim} \Delta l t)$ are drawn from an exponential distribution and the reaction with the smallest reaction time satisfying $\tau_k < \Delta t$ is executed.
If a reaction occurs, the motor starts at a random position within the chord length $\Delta l$. 
If no reaction occurs, the motor protein performs free diffusion, which is implemented in the same way as for free tubulin units.  
The dynamics of filament-bound motors is implemented as discussed above for single filaments.

\begin{figure}[htb]
\centering
\includegraphics{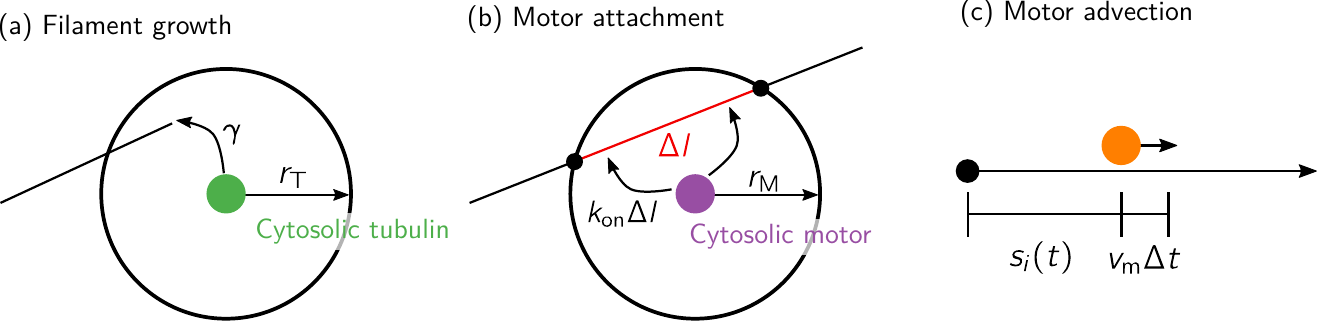}
\caption{%
(a) Schematic representation for filament growth. 
If a cytosolic tubulin unit is within a distance $r_\mathrm{T}$ of a filament plus end it detaches at rate $\gamma$. 
(b) Analogously, cytosolic motors attach to filaments at the rate $\kon \Delta l$ when they are at a distance $r_\mathrm{M}$ from the filament. Attachment occurs at a random position on the chord of length $\Delta l$. 
(c) Filament-bound motors move with constant velocity $\vm$ towards the filament plus end. 
}
\label{fig:sim-sketch}
\end{figure}

\section{Comparison to experimental data}

We choose the setup of our agent-based model comparable to the experimental system in Rank et al.~\cite{Rank2018}.
The experiments in ~\cite{Rank2018} were performed in a setup where GMP-CPP stabilized
microtubules were pre-grown at a concentration of $\barRhoT = \SI{2}{\micro M}$ in a confined system of \SI{100}{\micro l}. By varying the amount of time the microtubules were pre-grown the authors controlled the length distribution of microtubules at the time point of incubation with Kip3. After the initial microtubule growth period the solution was divided into compartments of \SI{25}{\micro l} and each compartment was supplemented with a dilution of Kip3 ranging from 0-\SI{400}{\nano M}. After \SI{60}{\min} of incubation with Kip3 the Kip3-MT interaction was terminated and microfluidic flow channels
were constructed to image the microtubules (see Supplemental Material of ~\cite{Rank2018} for details of the experimental setup). 

In experiments it is hard to control the precise level of protein concentrations. For example, the concentration of ``active'' Kip3 could not be determined in \cite{Rank2018} since the inactivation of Kip3 during the purification, snap-freezing and thawing process could not be quantified. In addition the authors expect Kip3 to form clusters, which would cause an additional amount of Kip3 that does not participate in the length regulation dynamics in particular for high motor concentrations.
In addition to the uncertainties in the protein concentrations, there are measurement inaccuracies because, for example, short microtubules could not be distinguished from tubulin clusters in experiaments.
When comparing experiments and agent based simulations we tried to account for this lack of resolution by only taking filament length $l_i > \SI{0.5}{\micro \m}$ into account. 
In particular as the protein concentrations that are actively involved in the length regulation process are not known in experiments a quantitative comparison between the experimental data and our agent based simulation is not possible. It is however possible to compare the results on a qualitative level.     

\begin{figure}[!tb]
\centering
\includegraphics[width=14cm]{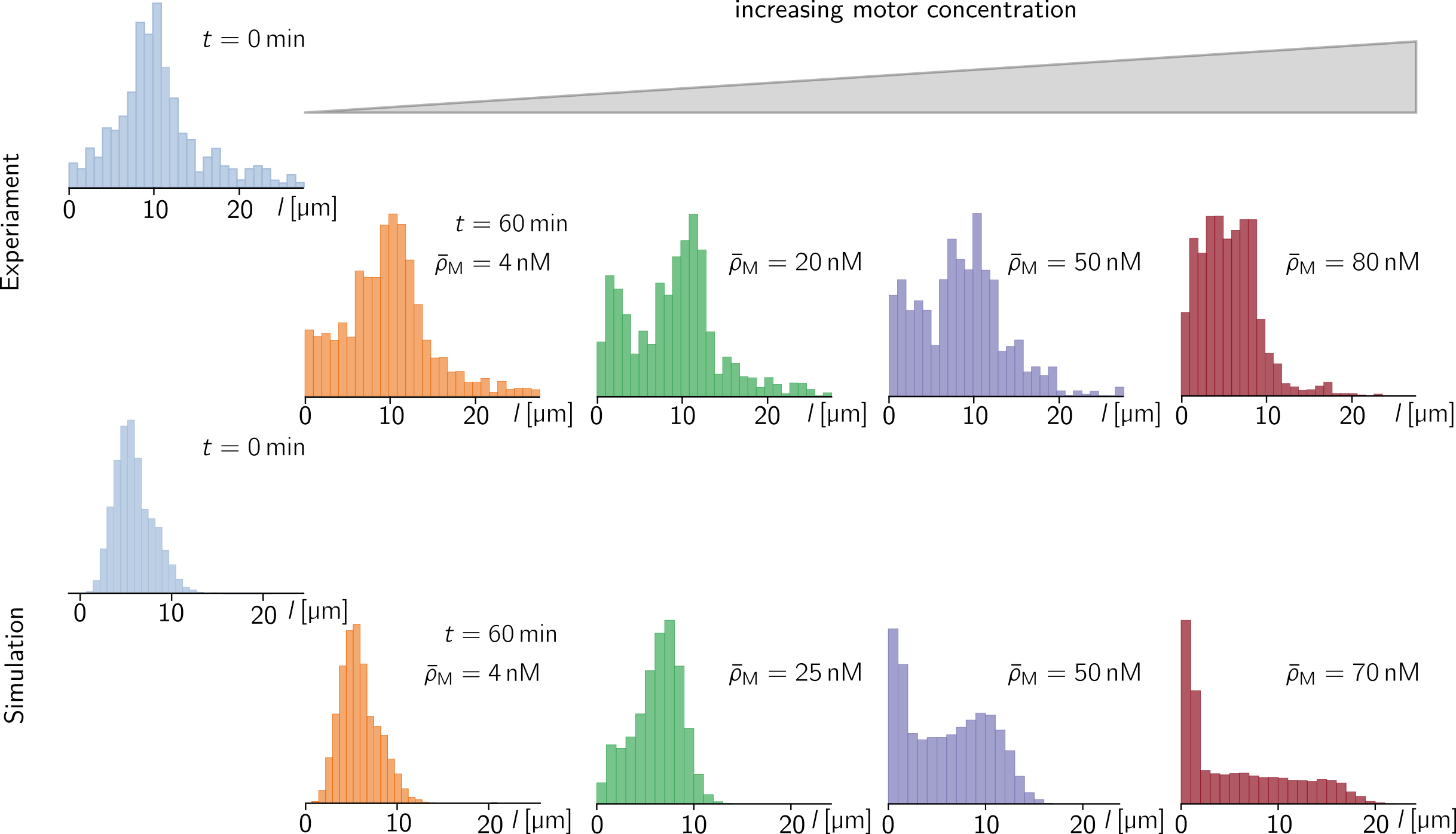}
\caption{%
    Measured filament length distributions after one hour of incubation at various concentrations of Kip3 indicated in the graph. The upper panel shows experimental data taken from \cite{Rank2018} and the lower panel shows results obtained from our agent-based simulations. The simulation parameters can be found in table~\ref{tab:parameters}
}
\label{fig:exp-comp}
\end{figure}

In our agent-based simulations we observe a slow initial dynamics consistent with the theoretical result of the linear stability analysis that predicts an initial growth rate of a perturbation which is on the order $\sim \mathcal{O}(10^{-4}\si{\s^{-1}})$. Once the initial perturbation has grown nonlinear effects start playing a role and the dynamics speeds up significantly. Our agent based simulations show that the system has not yet reached its steady state after one hour of incubation with Kip3, it is rather in the initial phase of pattern formation (cf. Fig. 4 in the main text). The measured length distribution after \SI{60}{\min} of incubation with Kip3 therefor strongly depend on the initial condition (initial length distribution). 
In \cite{Rank2018} the authors found that for narrow initial length distribution the filament length distributions did not change significantly from the initial length distribution after \SI{60}{\min} of incubation with Kip3 (cf. Fig. S20 in \cite{Rank2018}). Those observations are consistent with our theory and observations from the agent based simulation as the initial dynamics is slow (in particular for low motor concentrations) as stated above. 
For broader initial length distributions the authors observed different behaviour at different concentrations of Kip3 (see upper panel in Fig.~\ref{fig:exp-comp}). 
At low motor concentrations no significant change in the length distribution was observed. At intermediate concentrations the measured length distribution was bi-modal and at high motor concentrations the microtubule length distribution was broad ($\sim$ uniform). 
Those observations are consistent with measurements from our agent-based simulation (see lower panel in Fig.~\ref{fig:exp-comp}).
Note the volume per microtubule $V_0$ was estimated to be \SI{1.66}{\micro \m^3} in \cite{Rank2018}. 
Here however we used $V_0 = \SI{0.5}{\micro \m^3}$ resulting in less resources per filament and therefor shorter microtubules (this was done for the shake of computational feasibility). We used slightly higher tubulin concentrations ($\rhoT \sim \SI{2.5}{\micro M}$) in the agent based simulation to obtain similar length distributions. Moreover we find a comparable evolution of the measured microtubule length distribution; see Fig.~\ref{fig:exp-comp-time}.  

\begin{figure}[!tb]
\centering
\includegraphics{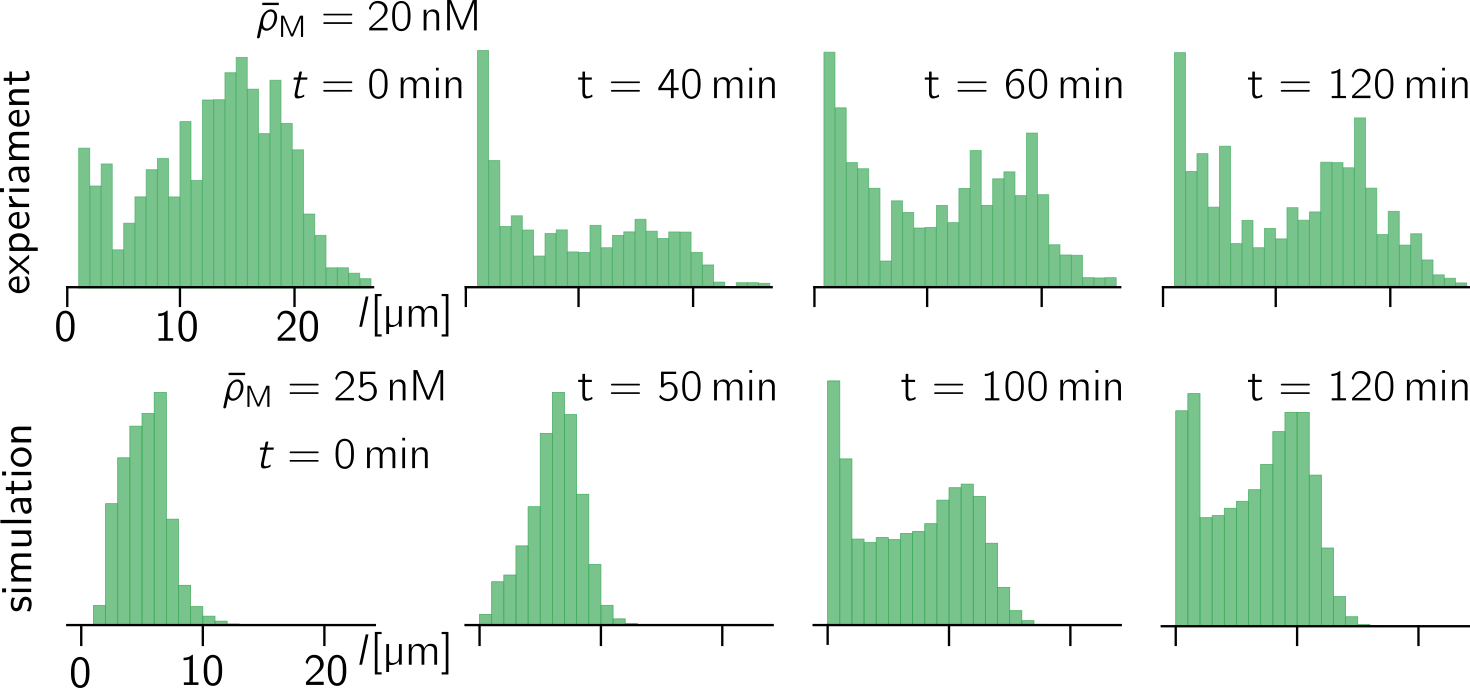}
\caption{%
Measured filament length distribution at fixed Kip3 concentrations at different time points. The upper panel shows experimental data taken from \cite{Rank2018} and the lower panel shows results obtained from our agent-based simulations. Simulation parameters are as specified in table~\ref{tab:parameters}.
}
\label{fig:exp-comp-time}
\end{figure}

Our agent-based simulations shows that the microtubule length distribution for narrow initial distributions and low motor concentrations also becomes broad but at significantly later times ($t \gg \SI{60}{\min}$). Moreover our simulations suggest that the bi-modal distribution observed in experiments for intermediate motor concentrations is just a transient phenomenon. At later times we observe (in the agent-based simulation) that the length distributions for both intermediate and low motor concentrations becomes more reminiscent of what is observed for high motor concentrations (as long as the concentrations are chosen in a range where patterns form see Fig.~\ref{fig:LSA-rhoM-rhoT-diagrams}(d)). 

We are recognizing that the experimental data available are not sufficient to fully validate our model.
A first experimental approach to validate our model would be to repeat the experiments described in \cite{Rank2018} and to measure the length distributions over a longer period of time to verify the predictions by our agent based simulations, namely that the length distributions at all motor concentrations were pattern formation is observed become similar. 
To gain further progress in understanding spatiotemporal dynamics, experiments that spatially and temporally resolve tubulin density would be desirable.

\section{Point-like filament approximation}

In this section, we provide a detailed analysis of the dynamics in the point-like filament approximation.
For the reader's convenience, we repeat the equations that govern the dynamics in this approximation
\begin{subequations}
\label{eq:point-like-dynamics}
\begin{align} 
    \partial_{t}l(\textbf{x},t) 
    &=  
    a\gamma \cT(\textbf{x},t) - \vsqss(\textbf{x},t)  \, ,\\
    \partial_{t} \cT(\textbf{x},t) 
     &= 
    \DT \partial_{\textbf{x}}^{2}\cT(\textbf{x},t) - \left[ \gamma \cT(\textbf{x},t) - \vsqss(\textbf{x},t)/a \right] V_{0}^{-1} \, ,\\
     \partial_t \rhoM (\textbf{x},t)
    &=
    \DM \partial_\textbf{x}^2 \cMqss (\textbf{x},t) \, ,
\end{align}
with
\begin{align}
    \cMqss(l,\rhoM) &= \frac{
        \rhoM
    }{
        1 + (l/l_\mathrm{c})^2
    } \, , \\
    \vsqss(l, \rhoM)
    &=
    a \kon l \cMqss(l, \rhoM) \, ,
\end{align}
\end{subequations}
where $l_\mathrm{c}^2 = 2 \vm V_0/\kon$.
These dynamics conserve the total numbers of tubulin units and motors
\begin{align}
    N_\mathrm{T} = \barRhoT L_x L_y &= \int_0^{L_x} \mathrm{d} x \int_0^{L_y} \mathrm{d} y \, \left[ \cT(\textbf{x},t) + \frac{l(\textbf{x},t)}{a V_0} \right] , \\
    N_\mathrm{T} = \barRhoM L_x L_y &= \int_0^{L_x} \mathrm{d} x \int_0^{L_y} \mathrm{d} y \, \rhoM(\textbf{x},t) .
\end{align}
The corresponding average densities $\barRhoT$ and $\barRhoM$ will be the main control parameters of interest in the following. 
We will also discuss the role of the diffusion constants $\DT$ and $\DM$.

We will first consider the homogeneous steady states, which are fixed points of the single-filament dynamics. 
In particular, our analysis will show that these fixed points can be read off from a graphical construction in the $l$--$\cT$ phase plane and that there is a regime of bistability.
Next, we will perform a linear stability analysis of the homogeneous steady states against spatial perturbations and discuss various limiting regimes and the role of the diffusion constants $\DT$ and $\DM$.
Finally, we will present numerical simulations of the equations that govern the dynamics in the point-like filament approximation. 
These simulations capture the patterns that initially emerge from the instability predicted from linear analysis. 
However, sharp gradients, on scales shorter than the filament lengths, rapidly emerge such that the underlying assumption of point-like filaments is violated.

\subsection{Homogeneous steady states}

The homogeneous steady states are given by the fixed points of the polymerization kinetics determined by the local balance of polymerization and depolymerization and conservation of the total density of tubulin
\begin{subequations} \label{eq:tubulin-fixed-point}
\begin{align}
    a\gamma \cT^* = \vsqss(l^*, \barRhoM)  \, , \\
    \cT^* + \frac{l^*}{a V_0} = \barRhoT \,.
\end{align}
\end{subequations}
In the $l$--$\cT$ plane, the solutions to these equations are given by the intersections between the \emph{nullcline} $\cT = \vsqss(l, \barRhoM)/(a \gamma)$ and the \emph{reactive phase space} $\cT = \barRhoT - l/(a V_0)$; see Fig.~\ref{fig:LSA-rhoM-rhoT-diagrams}(a,b).
This graphical construction allows us to gain insight into the behaviour of the homogeneous steady states as a function of parameters.
As long as the slope of the nullcline is larger than $-1/(a V_0)$ for all $l$, there is only a single fixed point [Fig.~\ref{fig:LSA-rhoM-rhoT-diagrams}(a)].
If there is a section where the nullcline slope is more negative than $-1/(a V_0)$, i.e., $\partial_l \vsqss < -\gamma/V_0$,
the reactive phase space can intersect the nullcline three times [Fig.~\ref{fig:LSA-rhoM-rhoT-diagrams}(b), case (iii)], giving three homogeneous steady states.
The bistable region is delimited by saddle-node (also called fold or limit point) bifurcations where the reactive phase space is tangential to the nullcline, i.e.\ where the nullcline slope is $\partial_l \vsqss = -\gamma/V_0$. 

To determine the stability against these steady states against spatially homogeneous perturbations, we linearize Eq.~(\ref{eq:point-like-dynamics}a) under the mass conservation constraint $\cT + l/(a V_0) = \barRhoT$ which yields $\partial_t \delta l = \sigma_\mathrm{poly} \delta l$, with $\sigma_\mathrm{poly} = \gamma/V_0 + \partial_l$.
Thus, the steady state that lies on the nullcline section with $\partial_l \vsqss < -\gamma/V_0$ is unstable against spatially homogeneous perturbations (empty disk) while the other two are stable (filled disks).

\begin{figure*}[!tb]
\centering
\hspace*{-1.5cm}\includegraphics{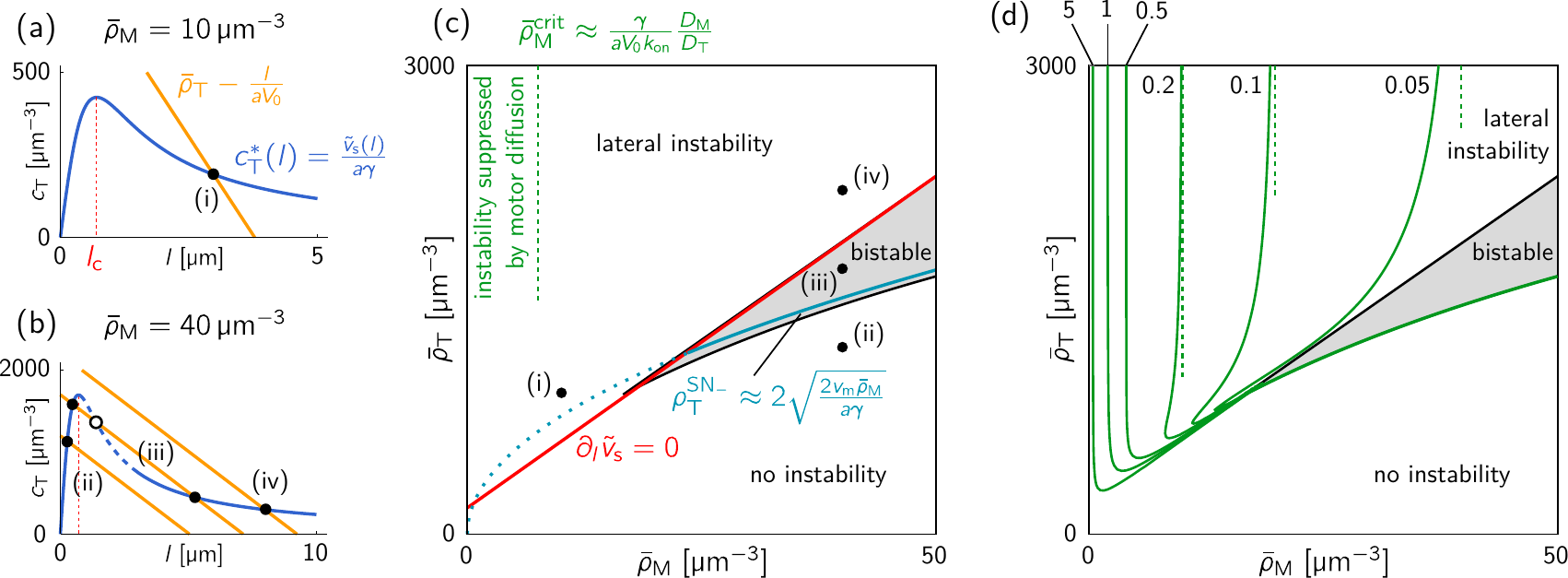}
    \caption{%
    Homogeneous steady states and linear stability analysis for the point-like filament approximation, Eqs.~(\ref{eq:point-like-dynamics}a)--(\ref{eq:point-like-dynamics}e). 
    (a), (b) $l$--$\cT$ phase plane analysis showing the graphical construction of the homogeneous steady states in the monostable case (a) and the bistable case (b). Homogeneous steady states are fixed points (black disks) of the length-regulation dynamics, found as intersection points of the nullcline $\cT^*(l) = \vsqss(l)/(a \gamma)$ (blue line) and the mass-conservation constraint for tubulin (orange line), $\cT = \barRhoT - l/(a V_0)$. The red dashed line marks the apex of the nullcline at $l_\mathrm{c}$.
    (c) Stability diagram in the $\barRhoM$--$\barRhoT$ plane. Black points labelled (i)--(iv) correspond to the scenarios shown in (a) and (b). The bistable regime (shaded in gray) is delimited by limit point bifurcations (black lines). The red line, given by Eq.~\eqref{eq:lc-line}, separates the regions of positive and negative nullcline slope. Negative nullcline slope is a necessary condition for lateral instability. In addition lateral instability requires sufficiently large motor density $\barRhoM > \barRhoT^\mathrm{crit}$, which is set by the diffusivity ratio $\DT/\DM$ [see panel (d)]. Below this threshold, instability is suppressed by motor diffusion as explained in the main text.
    (d) Contour lines of the critical diffusivity ratio $\DT/\DM$ above which the system is laterally unstable [cf.\ Eq.~\eqref{eq:diff-stability-threshold}]. For large $\barRhoT$, this ratio becomes independent of $\barRhoT$ and sets a threshold for the minimal motor density $\barRhoT^\mathrm{crit} \propto \DM/\DT$ as indicated by the dashed green lines [cf.\ panel (b)].
    }
\label{fig:LSA-rhoM-rhoT-diagrams}
\end{figure*}

The locations of these limit point bifurcations can be estimated by simple approximations, which highlight the role of the various parameters for the location of the bistable regime in the $\barRhoM$--$\barRhoT$ diagram.
Near the apex of the nullcline at $l = l_\mathrm{c}$ [marked by a red, dashed line in Fig.~\ref{fig:LSA-rhoM-rhoT-diagrams}(a,b)], its curvature is large such the slope $\partial_l \vsqss$ reaches $-\gamma/V_0$ near the apex.
Substituting $l^* = l_\mathrm{c}$ into the total density of tubulin and using $a \gamma \cT^* = \vsqss(l^*)$ yields
\begin{equation} \label{eq:lc-line}
    \barRhoT^\mathrm{apex} 
    = \frac{l_\mathrm{c}}{a V_0} + \frac{\vsqss(l_\mathrm{c})}{a \gamma}
    = \sqrt{\frac{2 v_\mathrm{m} V_0}{\kon}} \left(\frac{1}{a V_0} + \frac{\kon}{2 \gamma} \barRhoM\right).
\end{equation}
This shows that there is a linear relation between the motor density $\barRhoM$ and tubulin density at the apex, $\barRhoT^\mathrm{apex}$ [red line Fig.~\ref{fig:LSA-rhoM-rhoT-diagrams}(c)]. Note that this provides a good approximation of the upper edge of the bistable regime.

To estimate lower edge of the bistable regime, we approximate consider the limit $l^* \gg l_\mathrm{c}$. To lowest order, this yields $l^* \approx a V_0 \barRhoT$ and $\partial_l \vsqss \approx a \kon l_\mathrm{c}/l^2$. These approximations, substituted in to the criterion $\partial_l \vsqss = -\gamma/V_0$ yield
\begin{equation} \label{eq:saddle-node-estimate}
    \barRhoT^{\mathrm{SN}_-} \approx 2\sqrt{\frac{2 \vm \barRhoM}{a \gamma}},
\end{equation}
which is plotted as a teal line in Fig.~\ref{fig:LSA-rhoM-rhoT-diagrams}(c).

Together, the estimates Eq.~\eqref{eq:lc-line} and Eq.~\eqref{eq:saddle-node-estimate} reveal how the boundaries of the bistable regime depend on the system parameters. Increasing the polymerization rate $\gamma$ moves the bistable regime to lower tubulin densities and reduces its size while increasing the motor velocity $\vm$ move the bistable regime to higher tubulin densities. The attachment rate $\kon$ and volume per filament $V_0$ only affect the upper boundary of the bistable regime, moving it to higher tubulin densities when increased.

\subsection{Linear stability analysis} \label{sec:LSA}

Let us now analyse the stability of the homogeneous steady states against spatial perturbations. For a more compact notation, we combine the field variables into the vector $\mathbf{u} = (l, \cT, \rhoM)$. 
The homogeneous steady states are then $\mathbf{u}^* = (l^*, \cT^*, \barRhoM)$.
Writing the spatial perturbations in terms of Fourier modes
\begin{equation}
    \mathbf{u}(x,t) = \mathbf{u}^* + \delta \mathbf{u}_q e^{i \textbf{q} \cdot \textbf{x} + \sigma t} \;,  
\end{equation}
where $\textbf{q}$ denotes the wave vector, and linearizing Eqs.~\eqref{eq:point-like-dynamics} for small $\delta \mathbf{u}_q$ yields an eigenvalue problem for the growth rates $\sigma(q)$
\begin{equation}
    J(q) \delta \mathbf{u}_q = \sigma(q) \delta \mathbf{u}_q
\end{equation}
with the Jacobian (or stability matrix)
\begin{equation} \label{eq:full-Jacobian}
    J(q) = \begin{pmatrix}
        - \partial_l \vsqss & a \gamma & -\partial_{\rhoM} \vsqss \\
        \partial_l \vsqss/(a V_0) & - \gamma/V_0 - \DT q^2 & \partial_{\rhoM} \vsqss/(a V_0) \\
        -\DM q^2 \partial_l \cMqss & 0 & -\DM q^2 \partial_{\rhoM} \cMqss 
    \end{pmatrix}_{\!\mathbf{u}^*}.
\end{equation} 
In the following, we omit the subscript $\mathbf{u}^*$, with the understanding that all expressions are evaluated in the homogeneous steady state.

For spatially uniform perturbations, $q = 0$, the last row of the Jacobian vanishes and the eigenvalues of $J(0)$ are given by
\begin{equation}
    \sigma_1 
    = 
    \sigma_2 
    = 0, 
    \quad 
    \sigma_3 
    = - \partial_l \vsqss - \frac{\gamma}{V_0}
    \;.
\end{equation}
The two zero eigenvalues correspond to perturbations that would change the average total densities $\barRhoT$ and $\barRhoM$, respectively. 
In a closed system with conservation of mass, however, these disturbances are not possible.
Thus, the remaining eigenvalue $\sigma_3$ determines the stability to homogeneous perturbations that respect conservation of mass. 
One finds an instability for $\partial_l \vsqss < - \frac{\gamma}{V_0}$, which confirms the geometric criterion (in the in the $l$--$\cT$ phase space) discussed above.

In general, the stability of a spatially uniform steady state against spatial perturbations with a wave number $q$ is determined by the dispersion relation $\sigma(q)$; a typical dispersion relation in the laterally unstable parameter regime is shown in Fig.~3(a) in the main text.

While the eigenvalues of the full $3{\times}3$ Jacobian Eq.~\eqref{eq:full-Jacobian} can in principle be found analytically, the resulting expressions are not insightful. We therefore determine the eigenvalues numerically.
Notably, we find that the instability is always of long-wavelength type, i.e., the band of unstable modes extends to $q \to 0$.
Next, we will further analyze this long-wavelength limit. 
Further below, we will also consider the limit of well-mixed tubulin. In both these limits the resulting effective Jacobians are $2{\times}2$ matrices which allows us to find closed expressions for the instability criteria.

\subsubsection{Long-wavelength limit}
Since we numerically find that instability onset is always at long wavelengths, i.e.\ in the limit $q \to 0$, we can determine the instability criterion by considering the limit of small wavenumbers $q$.
In this long-wavelength limit, we can make the local quasi-steady state approximation $l(x,t) \approx l^*(x,t)$, $\cT(x,t) \approx \cT^*(x,t)$, where the local equilibrium $(l^*, \cT^*)$ depends on the local total densities $\rhoT$, $\rhoM$.
The dynamics for the tubulin density is obtained by adding the equations for $l$ and $\cT$, Eqs.~\eqref{eq:point-like-dynamics}(a,b). Using the local quasi-steady state approximation then yields
\begin{subequations}
\begin{align} \label{eq:rhoT-dyn}
    \partial_t \rhoT &= \DT \nabla^2 \cT^*(\rhoT, \rhoM) \; , \\
    \partial_t \rhoM &= \DM \nabla^2 \cMqss[l^*(\rhoT,\rhoM), \rhoM] \; .
\end{align}
\end{subequations}
The respective long-wavelength (LW) Jacobian is given by
\begin{equation}
    J_\mathrm{LW} (q) = -q^2 \begin{pmatrix}
        \DT \partial_{\rhoT} \cT^* & \DT \partial_{\rhoM} \cT^* \\
        \DM \partial_{\rhoT} \cMqss & \DM \partial_{\rhoM} \cMqss
    \end{pmatrix}
    \; ,
\end{equation}
and its eigenvalues read (as for any $2\times2$ matrix)
\begin{equation} \label{eq:trace-det-relation}
    \sigma_{1,2} (q)
    = 
    \frac12 
    \bigg[
    \mathrm{tr} \,  J_\mathrm{LW} (q) \pm \sqrt{\big(\mathrm{tr} \,   J_\mathrm{LW} (q) \big)^2 - 4 \, \mathrm{det} \,   J_\mathrm{LW} (q)} \,
    \bigg] \; .
\end{equation}
Therefore, there is an instability (positive eigenvalue $\sigma_i$) either when the trace ($\sigma_1 + \sigma_2$) is positive or when the determinant ($\sigma_1 \cdot \sigma_2$) is negative.  
Moreover, when the determinant is positive, $\mathrm{det} \, J_\mathrm{LW} > 0$, the eigenvalues' real parts cross zero as a pair of complex conjugates when $\mathrm{tr} \, J_\mathrm{LW}$ crosses zero.
This indicates an oscillatory instability (Hopf bifurcation) where the oscillation frequency is determined by the imaginary part of the eigenvalues.
(Further away from the onset of instability, imaginary part vanishes for the fastest growing mode, so the instability loses it's oscillatory character; see Fig.~3(a) in the main text.)
$J_\mathrm{LW} (q)$ has a positive trace if
\begin{equation} \label{eq:stablity-criterion}
     \DT \partial_{\rhoT} \, \cT^* + \DM \, \partial_{\rhoM} \cMqss < 0 \; .
\end{equation}
We will discuss this instability criterion below. Since, $\mathrm{det} \, J_\mathrm{LW} > 0$, the instability will be oscillatory near onset, as noted above. 

A positive eigenvalue could also result from a negative determinant. However, numerically, we find that the determinant seems to be always positive, although we could not show this analytically.
There is also a physical reasoning why the determinant should always be positive. 
First, observe that the diffusion constants can be factored out. 
The remaining term, which then determines the sign of the determinant is independent of the diffusion constants. Therefore, if this term were negative, it would imply an instability independent of the diffusion constants. 
This is at odds with physical intuition for a diffusion driven lateral instability,  
in particular, because sufficiently fast motor diffusion will always act to suppress lateral instability.

Let us now analyze the stability criterion Eq.~\eqref{eq:stablity-criterion}.
First consider the case $\DM = 0$. 
Then the instability condition simply reads $\partial_{\rhoT} \cT^* < 0$, which can be  understood as follows. 
When $
\DM = 0$, the total motor density remains constant under the dynamics and therefore spatially uniform (by choice of initial condition). 
The only dynamic variable in the long-wavelength limit is then the tubulin density, governed by
\begin{equation} \label{eq:tubulin-effective-diffusion}
    \partial_t \rhoT = \DT \nabla^2 \cT^*(\rhoT, \barRhoM) = \nabla \big[ \DT \partial_{\rhoT} \cT^*(\rhoT, \barRhoM) \nabla \rhoT \big], 
\end{equation}
where we used the chain rule for the second equality. This is a diffusion equation with effective diffusion constant $\DT \partial_{\rhoT} \cT^*$. 
Thus, if $\partial_{\rhoT} \cT^*$ is negative, there is an instability corresponding to effective ``anti-diffusion.''
This is the basic mechanism underlying mass-redistribution instability in the long-wavelength limit \cite{Brauns2020}.
This shows that the basic mechanism of the instability is the same as in protein-based pattern forming systems like the ones discussed in \cite{Halatek2018}.

In the main text, we discuss the stability condition based on the nullcline slope $\partial_l \vsqss$. To relate the derivative $\partial_{\rhoT} \cT^*$ to the nullcline slope $\partial_l \vsqss$, we apply the derivative $\partial_{\rhoT}$ to the fixed point equations Eq.~\eqref{eq:tubulin-fixed-point} which yields 
\begin{equation} \label{eq:del_rhoT_cT}
    \partial_{\rhoT} \cT^* = \frac{\partial_l \vsqss}{\frac{\gamma}{V_0} + \partial_l \vsqss}.
\end{equation}
Thus, along sections of the nullcline where $\partial_l \vsqss > -\gamma/V_0$ (implying stability against homogeneous perturbations), the lateral instability condition $\partial_{\rhoT} \cT^* < 0$ is analogous to the slope condition $\partial_l \vsqss < 0$ discussed in the main text.
In the $\barRhoM$--$\barRhoT$ plane, the line $\barRhoT^\mathrm{apex}(\barRhoM)$, marks the nullcline apex where $\partial_l \vsqss = 0$ [red line in Fig.~\ref{fig:LSA-rhoM-rhoT-diagrams}(c)].
Above this line, the nullcline slope is negative, and therefore, the homogeneous steady state is laterally unstable.
In other words, a sufficiently high tubulin density compared to the motor density is required for the instability to occur.

In the bistable regime, we have to distinguish between the two stable steady states. Notably, the nullcline slope is always negative for the large-$l$ fixed point (see Fig.~\ref{fig:LSA-rhoM-rhoT-diagrams}(c). Therefore, this fixed point always laterally unstable.
In contrast the low-$l$ fixed point is unstable only in a very narrow regime due to the high nullcline curvature near its apex.
In the stability diagram in Fig.~3(b) in the main text, we show the stability of the large-$l$ fixed point. 

The above conditions for lateral instability are necessary and sufficient in the case $\DM = 0$. Let us now turn to the case $\DM > 0$.
As we heuristically argued in the main text, motor diffusion generally counteracts lateral instability.
Indeed, the stability threshold $\mathrm{tr}\,J_\mathrm{LW} = 0$ obtained from the above linear stability analysis gives
\begin{equation} \label{eq:trace-stability-threshold}
    \DT \partial_{\rhoT} \cT^* 
    = 
    -\DM \partial_{\rhoM} \cMqss 
    \; .
\end{equation}
Substituting Eq.~\eqref{eq:del_rhoT_cT} for $\partial_{\rhoT} \cT^*$ and using $\partial_{\rhoM} \cMqss = \cMqss/\barRhoM$ [cf. Eq.~(\ref{eq:point-like-dynamics}d)], we can write this condition as
\begin{equation} \label{eq:diff-stability-threshold}
    \frac{\DT}{\DM} = \frac{\cMqss}{\barRhoM} \frac{\frac{\gamma}{V_0} + \partial_l \vsqss|_{l^*}}{- \partial_l \vsqss |_{l^*}}
    \; ,
\end{equation}
determining the threshold value for the ratio of the diffusion constants $\DT/\DM$. 
For diffusivity ratios below this threshold, i.e.\ for too fast motor diffusion, the instability is suppressed [see red lines in Fig.~\ref{fig:DM-DT-stability-diagram}].
In Fig.~\ref{fig:LSA-rhoM-rhoT-diagrams}(d), contour lines show the instability threshold in the $\barRhoM$--$\barRhoT$ plane for several diffusivity ratios. 

Notably, for sufficiently large $\barRhoT$, the threshold becomes independent of $\barRhoT$.
The critical motor density $\barRhoM^\mathrm{crit}$ in this regime is indicated by the he dashed green lines in Fig.~\ref{fig:LSA-rhoM-rhoT-diagrams}(c,d). 
To obtain an estimate for this critical motor density $\DT/\DM$, we approximate Eq.~\eqref{eq:diff-stability-threshold} in the limit $l^* \gg l_\mathrm{c}$ and solve for $\barRhoM$:
\begin{equation} \label{eq:diff-threshold-approx}
    \barRhoM^\mathrm{crit} \approx \frac{\gamma}{a \kon V_0} \frac{\DM}{\DT}.
\end{equation}
This shows that the critical motor density is proportional to the diffusivity ratio $\DM/\DT$.
A higher motor diffusivity requires a higher motor density for the instability to occur.
At first glance, this may seem somewhat counterintuitive.
The reason for this effect is that increasing the motor density increases the magnitude of the nullcline slope since $\vsqss \propto \rhoM$. This, in turn, increases the growth rate of the tubulin-mass-redistribution instability [cf.\ Eq.~\eqref{eq:tubulin-effective-diffusion}], which allows it to overcome the stabilizing effect of motor diffusion. 
Physically, a steeper nullcline slope means that gradients in the filament length lead to steeper gradients in cytosolic tubulin concentration.

Substituting the values for $a, \gamma$, and $\kon$ from Table~\ref{tab:parameters} into Eq.~\eqref{eq:diff-stability-threshold} yields the condition for the number of motors per filament, $V_0 \barRhoM \gtrsim 0.57 \DM/\DT$, as given in the main text.
In the $\barRhoM$--$\barRhoT$ diagram, Eq.~\eqref{eq:diff-threshold-approx} sets an approximate lower bound for the unstable region [see dashed green lines in Fig.~\ref{fig:LSA-rhoM-rhoT-diagrams}(c,d)]. Below this threshold, the instability is suppressed by motor diffusion. In the limit $\DT \gg \DM$, the critical value $\barRhoM^\mathrm{crit}$ goes to zero. 
For physiological parameters and the estimated diffusivity ratio $\DT/\DM \approx 6$, the threshold is at about $V_0 \barRhoM^\mathrm{crit} \approx 0.1$ motors per filament.

\begin{figure}[tb]
\centering
\includegraphics{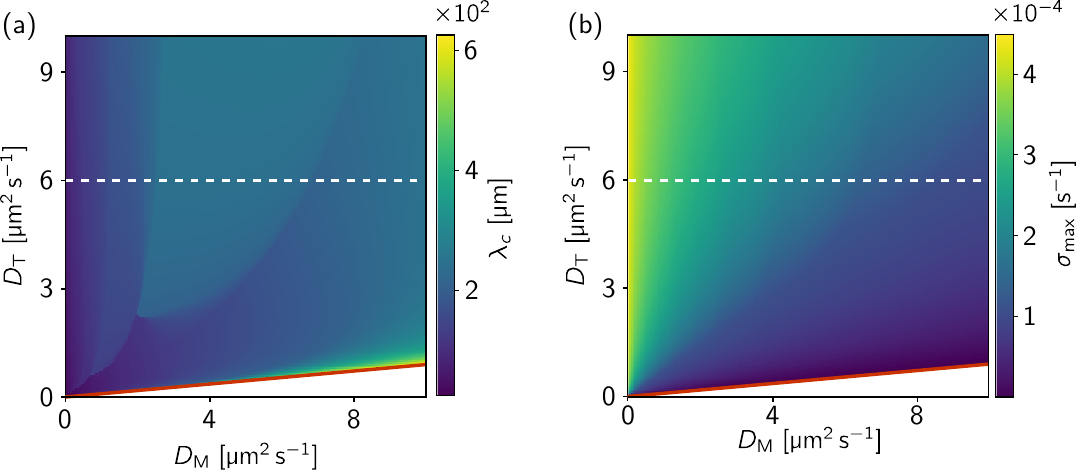}
    \caption{%
    (a) Wavelength and (b) growth rate of the fastest growing mode in as a function of the diffusion constants $\DM$ and $\DT$. The white dashed line indicates the physiological diffusion constant of tubulin. Remaining parameters are as in Fig.~4 in the main text. The boundary of the regime of instability is indicated by a red line, corresponding to the critical diffusivity ratio given by Eq.~\eqref{eq:diff-stability-threshold}.
    }
\label{fig:DM-DT-stability-diagram}
\end{figure}

\subsubsection{Well-mixed cytosolic tubulin}
Above, we have analyzed the long-wavelength limit, where the polymerization kinetics can be assumed to be in a local quasi-steady state.
We now turn to the dynamics at short wavelengths, where cytosolic diffusion of tubulin is faster than the polymerization kinetics.
Relaxation to the local steady state length happens at the rate $\sigma_\mathrm{poly} = \gamma/V_0 + \partial_l \vsqss$, which we derived in the stability analysis for homogeneous perturbations above.
The rate of diffusive transport for modes with wavenumber $q$ is given by $\DT q^2$. Thus, for wavenumbers $q \gg \sqrt{|\sigma_\mathrm{poly}| / \DT}$, the cytosolic tubulin density can be assumed well-mixed. 
(A detailed discussion of this ``\emph{reaction-limited regime}'' and the complementary ``\emph{diffusion-limited regime}'' at large wavelengths is given in Ref.~\cite{Brauns2020} in the context of mass-conserving two-component reaction--diffusion equations.)

For spatial non-uniform perturbations ($q \neq 0$), a well-mixed (WM) cytosolic tubulin implies $\delta \cT = 0$. Thus, the reduced Jacobian is obtained by removing the central row and column from the full Jacobian $J$, Eq.~\eqref{eq:full-Jacobian}, and reads 
\begin{equation}
    J_\mathrm{WM}(q) = \begin{pmatrix}
    -\partial_l \vsqss & -\partial_{\rhoM} \vsqss \\
    -D_\mathrm{M} q^2 \partial_l \cMqss & -D_\mathrm{M} q^2 \partial_{\rhoM} \cMqss
\end{pmatrix}.
\end{equation}
As above the eigenvalues of this $2 {\times} 2$ matrix can be obtained from its determinant and trace
\begin{align}
    \mathrm{det} \, J_\mathrm{WM}(q) &= D_\mathrm{M} q^2 \kon \cMqss^2/\rhoM > 0 \; , \\
    \mathrm{tr} \, J_\mathrm{WM}(q) &= -\partial_l \vsqss - D_\mathrm{M} q^2 \cMqss/\rhoM \; ,
\end{align}
where we used $\vsqss \propto \cMqss \propto \rhoM$ [cf.\ Eq.~(\ref{eq:point-like-dynamics}d,e)] such that $\partial_{\rhoM} \cMqss = \cMqss/\rhoM$ and $\partial_{\rhoM} \vsqss = \vsqss/\rhoM$.
Figure 3(a) in the main text shows that the dispersion relation derived $J_\mathrm{WM}(q)$ agrees well with the dispersion relation of the full Jacobian, Eq.~\eqref{eq:full-Jacobian}, for sufficiently large $q$. 
Since the $\mathrm{det} \, J_\mathrm{WM}(q)$ is always positive, $J_\mathrm{WM}(q)$ has an unstable eigenvalue if and only if $\mathrm{tr}\, J_\mathrm{WM}(q) > 0$.
Thus, if $-\partial_l \vsqss > 0$, there is a band of unstable modes $q \in (0, q_\mathrm{max})$ with
\begin{equation} \label{eq:q-max}
    q_\mathrm{max}^2 = - \frac{\partial_l \vsqss}{D_\mathrm{M}}
    \frac{\barRhoM}{\cMqss}.
\end{equation}
This shows that cytosolic motor diffusion suppresses a lateral instability on short length scales. 
In particular, the band of unstable modes vanishes in the limit $D_\mathrm{M} \rightarrow \infty$. 
Therefore, an approximation in which both the cytosolic tubulin and the motors are assumed to be well-mixed will not reproduce the instability. 

Equation~\eqref{eq:q-max}, derived under the assumption of well-mixed cytosolic tubulin, only approximates edge of unstable modes of the full Jacobian Eq.~\eqref{eq:full-Jacobian}. This approximation is valid if $q_\mathrm{max}^2 \DT \gg |\sigma_\mathrm{poly}|$. Substituting the expressions and rearranging the terms yields the condition
\begin{equation}
       \frac{\DT}{\DM} \gg \frac{\cMqss}{\barRhoM} \frac{\frac{\gamma}{V_0} + \partial_l \vsqss|_{l^*}}{- \partial_l \vsqss |_{l^*}} \;.
\end{equation}
Comparing to Eq.~\eqref{eq:diff-stability-threshold} shows that the approximation is valid deep in the unstable regime, far from the instability threshold.

\subsection{Finite-element simulations}

To complement linear stability analysis, we performed numerical simulations of the point-like filament dynamics, governed by Eqs.~\eqref{eq:point-like-dynamics}. Specifically, a finite element method implemented in the COMSOL Mulitphysics software was used (see \texttt{point-like-filament-PDEs.mph}). 
To ensure numerical stability, we add a small diffusion term (diffusion constant \SI{1e-2}{\micro m^2.s^{-1}}) to the dynamics of $l(\mathbf{x},t)$. 
This is necessary because sharp interfaces emerge rapidly around the long-filament clusters that from the initial instability (Fig.~\ref{fig:point-like-simulation}).
Because of these sharp gradients, the assumption of point-like filaments underlying these simulations is no longer valid. Still, it is instructive to discuss the dynamics in this regime as it shares some of the features with the agent-based simulations with spatially extended filaments.

\begin{figure}[b]
\centering
\includegraphics{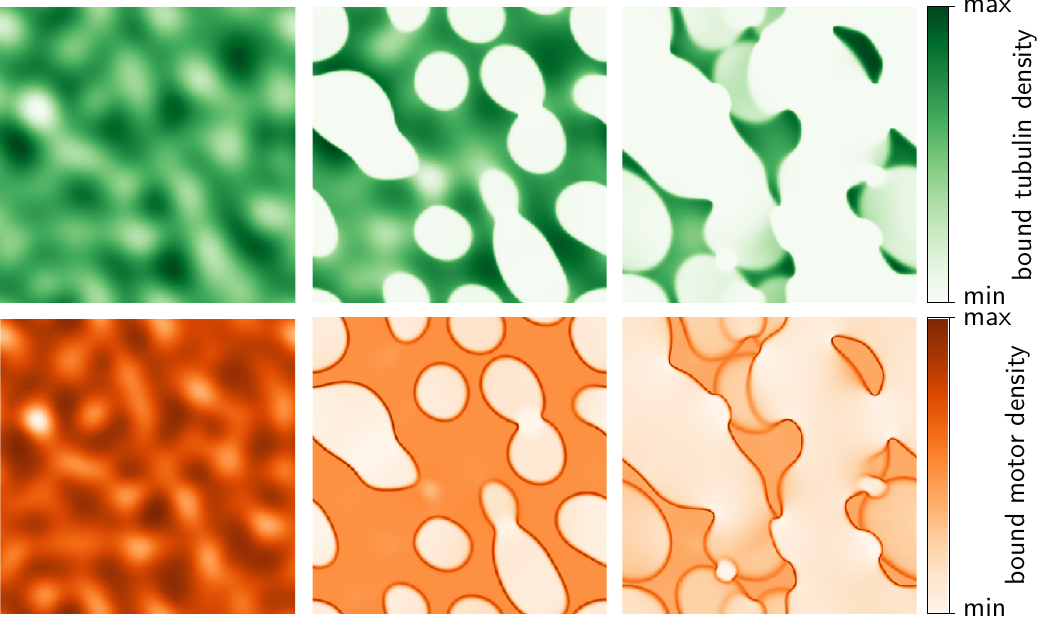}
    \caption{%
    Snapshots from finite-element simulations of Eqs.~(\ref{eq:point-like-dynamics}a)--(\ref{eq:point-like-dynamics}e) at times $1.8\times 10^{4}$, $2.5\times 10^{4}$, and $3.1\times 10^{4}$ from left to right.
    Top row: Filament length $l$ (bound tubulin density). The color bar ranges [min, max], are $[4, 7]$, $[0,15]$, $[0,32]$ from left to right.
    Bottom row: Filament-bound motor density $\tilde{M}$. The color scheme is logarithmic, with ranges $[10^{1.1},10^{1.2}]$, $[10^0,10^3]$, and $[10^0,10^3]$ from left to right. 
    Note that the concentration of bound motors is highest around the perimeter of regions with long filaments (high bound tubulin density).
    Domain size $L_x = L_y = \SI{500}{\micro m}$. Remaining parameters as in Figs.~4 and 5 in the main text.
    }
\label{fig:point-like-simulation}
\end{figure}

The interfaces of clusters that emerge from the initial instability propagate such that the regions of long filaments become smaller. This propagation is driven by the diffusive influx of motors from the regions where filaments are short, and therefore, most motors are in the cytosol. The motors diffusive into the long-filament regions, where they rapidly attach near the interface. This drives depolymerization of filaments near the interface, causing the long-filament regions to shrink.
The released tubulin units then diffuse in the cytosol and drive further growth of filaments in the long-filament regions.

In fact, the propagation of interfaces is already indicated by the dispersion relation. As we discussed below Eq.~\eqref{eq:trace-det-relation}, a positive determinant of $J_\mathrm{WM}$ implies that the eigenvalues are a pair of complex conjugates near the onset of instability. In the dispersion relation, this means that $\sigma(q)$ has a non-zero imaginary part near the zero crossing of its real part at $q = q_\mathrm{max}$ [see Fig.~3(a) in the main text].
In a previous study on mass-conserving reaction diffusion systems, we found that the properties of interfaces can be inferred from the right edge of the dispersion relation \cite{Brauns2020}. Specially, non-zero imaginary part at $q_\mathrm{max}$ indicates that the mode that defines the interface is propagating. 

In the point-like filament approximation, interfaces will continue to propagate until the long-filament clusters have completely disappeared. Subsequently, new clusters will emerge from lateral instability. And these clusters will again collapse due to interface propagation, driven by diffusive flux of motors into the clusters. 
In contrast, in agent-based simulations with spatially extended filaments, advective transport of motors along filaments counteracts diffusive influx of motors into the clusters. 
This is because the net orientation of the filaments is aligned with the density gradients and leads out of the clusters.
As a result of this advective motor transport, interface motion arrests eventually, thus producing the final, aster-like steady state structure (see Movie~1 and Fig.~4 in the main text).

\vspace*{1cm}
\begin{algorithm}[H]
	\caption{Approximative simulation scheme}
	\begin{algorithmic}[1]
		\While {$t< t_\mathrm{max}$}
		    \For{motor position $= \{s_1,s_2,\ldots\}$}
		        \State $s_i \rightarrow s_{i} + \vm\Delta t$
		        \If{$s_i > l$}
		            \State{\# cytosolic motors += 1}
		            \State{\# cytocolic tubulin += 1}
		            \State{filament length ${-}{=}$ $a$}
		        \EndIf
		    \EndFor
		    \State $r = $random\_real $\in (0,1)$
		    \If{$r < \kon l \Delta t \; (\# \text{cytosolic motors})$}
		        \State $s_k = $random\_real $\in (0,l)$
		        \State motor\_positions.append$(s_k)$
		        \State{\# cytosolic motors ${-}{=}$ 1}
		    \EndIf
		    \State $r =$ random\_real $\in (0,1)$
		    \If{$r < \gamma \Delta t \; (\# \text{cytosolic tubulin})$}
		    \State $l$ ${+}{=}$ $a$
		    \State{\# cytosolic tubulin ${-}{=}$ 1}
		    \EndIf
		\EndWhile
	\end{algorithmic} 
\end{algorithm}  

\begin{algorithm}[H]
	\caption{Filament growth dynamics}\label{alg:fil-growth}
	\begin{algorithmic}[1]
	    \For{position $(x_i,y_i)$ in free tubulin positions = $\{(x_0,y_0),\ldots \}$}
	    \State reaction time $\tau_{k} = \Delta t$
        \State filament plus ends in range = EuclidianDistance[$(x_i,y_i)$, plus\_end\_position] $< r_\mathrm{T}$
	    \For{filament plus end \textbf{in} filament plus ends in range}
	        \State $\tau_k^\mathrm{new} \rightarrow \gamma\exp(\gamma t)$
	        \If{$\tau_k^\mathrm{new} < \tau_{k}$}
	            \State $\tau_k = \tau_k^\mathrm{new}$
	            \State{remember filament plus end} 
	        \EndIf
	    \EndFor
	    \If{remembered filament plus end ${!}{=}$ empty}
	        \State grow filament
	        \State remove tubulin position from free tubulin positions
	    \Else
	        \State $x_i$ ${+}{=}$ $\sqrt{24 \DT \Delta t} \, \cdot  (\text{random\_real} \in (-0.5,0.5))$
	        \State $y_i$ ${+}{=}$ $\sqrt{24 \DT \Delta t} \, \cdot  (\text{random\_real} \in (-0.5,0.5))$
	    \EndIf
	    \EndFor
	\end{algorithmic} 
\end{algorithm}

\vspace*{1cm}

\begin{algorithm}[H]
	\caption{Free motor dynamics}
	\begin{algorithmic}[1]
	    \For{position $(x_i,y_i)$ in free motor positions = $\{(x_0,y_0),\ldots \}$}
	    \State reaction time $\tau_{k} = \Delta t$
        \State filaments in range = all filaments that intersect with a circle of radius 
        \State$r_\mathrm{M}$ around  $(x_i,y_i)$ (see Fig.~\ref{fig:sim-sketch}) 
	    \For{filament \textbf{in} filaments in range}
	        \State calculate $\Delta l$
	        \State $\tau_k^\mathrm{new} \rightarrow \kon \Delta l\exp(\kon \Delta l t)$
	        \If{$\tau_k^\mathrm{new} < \tau_{k}$}
	            \State $\tau_k = \tau_k^\mathrm{new}$
	            \State{remember filament} 
	        \EndIf
	    \EndFor
	    \If{remembered filament $!=$ empty}
	        \State attach motor to remembered filament at random position within $\Delta l$
	        \State set motor state to filament-bound
	    \Else
	        \State $x_i$ ${+}{=}$ $\sqrt{24 \DM \Delta t} \, \cdot  (\text{random\_real} \in (-0.5,0.5)$)
	        \State $y_i$ ${+}{=}$ $\sqrt{24 \DM \Delta t} \, \cdot (\text{random\_real} \in (-0.5,0.5)$)
	    \EndIf
	    \EndFor
	\end{algorithmic} 
\end{algorithm}

\clearpage

\end{widetext}

%

\end{document}